\documentclass[11pt,a4paper]{article}

\usepackage{graphicx,amsfonts,amsbsy, amssymb, amsthm, amsmath, mathrsfs}
\usepackage{manfnt}
\usepackage{tikz}                       % For the pictures
\usepackage{pgfplots}                   % For the plots

\pgfplotsset{%    Global settings for plots; can be overridden...
  height=8cm,
  width=10.5cm,
  legend pos=south east}

\usepackage[text={15cm,23cm}]{geometry} % Imposta pagina

\usepackage[noblocks]{authblk}          % For the authors

\renewcommand{\vec}[1]{{\bf{#1}}}

\renewcommand{\Re}{\mathop{\mathfrak{Re}}}

\newcommand{\rmd}{{\mathrm d}}
\newcommand{\rme}{{\mathrm e}}
\newcommand{\rmi}{{\mathrm i}}

\newcommand{\Ord}{{\mathrm O}}

\DeclareSymbolFont{lettersA}{U}{pxmia}{m}{it}

\DeclareMathSymbol{\alphaup}{\mathord}{lettersA}{"0B}
\DeclareMathSymbol{\betaup}{\mathord}{lettersA}{"0C}
\DeclareMathSymbol{\gammaup}{\mathord}{lettersA}{"0D}
\DeclareMathSymbol{\deltaup}{\mathord}{lettersA}{"0E}
\DeclareMathSymbol{\epsilonup}{\mathord}{lettersA}{"22}
\DeclareMathSymbol{\zetaup}{\mathord}{lettersA}{"10}
\DeclareMathSymbol{\etaup}{\mathord}{lettersA}{"11}
\DeclareMathSymbol{\thetaup}{\mathord}{lettersA}{"12}
\DeclareMathSymbol{\iotaup}{\mathord}{lettersA}{"13}
\DeclareMathSymbol{\kappaup}{\mathord}{lettersA}{"14}
\DeclareMathSymbol{\lambdaup}{\mathord}{lettersA}{"15}
\DeclareMathSymbol{\muup}{\mathord}{lettersA}{"16}
\DeclareMathSymbol{\nuup}{\mathord}{lettersA}{"17}
\DeclareMathSymbol{\xiup}{\mathord}{lettersA}{"18}
\DeclareMathSymbol{\piup}{\mathord}{lettersA}{"19}
\DeclareMathSymbol{\rhoup}{\mathord}{lettersA}{"1A}
\DeclareMathSymbol{\sigmaup}{\mathord}{lettersA}{"1B}
\DeclareMathSymbol{\tauup}{\mathord}{lettersA}{"1C}
\DeclareMathSymbol{\upsilonup}{\mathord}{lettersA}{"1D}
\DeclareMathSymbol{\phiup}{\mathord}{lettersA}{"1E}
\DeclareMathSymbol{\chiup}{\mathord}{lettersA}{"1F}
\DeclareMathSymbol{\psiup}{\mathord}{lettersA}{"20}
\DeclareMathSymbol{\omegaup}{\mathord}{lettersA}{"21}

\newcommand{\vecalpha}{{\pmb{\alphaup}}}

\newcommand{\vecgamma}{{\pmb{\gammaup}}}

\newcommand{\vectheta}{{\pmb{\thetaup}}}

\newcommand{\vecpsi}{{\pmb{\psiup}}}

\renewcommand{\Psi}{\varPsi}
\renewcommand{\Lambda}{\varLambda}
\renewcommand{\Sigma}{\varSigma}
\renewcommand{\Gamma}{\varGamma}
\renewcommand{\Theta}{\varTheta}
\renewcommand{\Xi}{\varXi}
\renewcommand{\Pi}{\varPi}
\renewcommand{\Upsilon}{\varUpsilon}
\renewcommand{\Phi}{\varPhi}
\renewcommand{\Omega}{\varOmega}

\newcommand{\Q}{{\mathbb Q}}
\newcommand{\Z}{{\mathbb Z}}

\newcommand{\Prob}{{\mathbb P}}
\newcommand{\E}{{\mathbb E}}

\newcommand{\tr}{\mathop{\rm Tr}}
\newcommand{\id}{\mathop{\rm id}\nolimits}

\newcommand{\cosec}{\mathop{\rm cosec}\nolimits}

\newcommand{\coloneq}{\mathbin{\hbox{\raise0.08ex\hbox{\rm :}}\!\!=}}
\newcommand{\eqcolon}{\mathbin{=\!\!\hbox{\raise0.08ex\hbox{\rm :}}}}

\renewcommand{\leq}{\leqslant}
\renewcommand{\geq}{\geqslant}
\renewcommand{\epsilon}{\varepsilon} % I prefer a more curly epsilon
\newcommand{\I}{{1\!\!1}}

\newcommand \printdate[3]{%
    \def \@suffix##1{%
        \def \@n{##1}%
        \ifnum \@n = 1 st\else%
        \ifnum \@n = 2 nd\else%
        \ifnum \@n = 3 rd\else%
        \ifnum \@n = 21 st\else%
        \ifnum \@n = 22 nd\else%
        \ifnum \@n = 23 rd\else%
        \ifnum \@n = 31 st\else%
        th\fi \fi \fi \fi \fi \fi \fi%
    }%
    \relax%
    \number #1\raise0.7ex\hbox{\footnotesize \@suffix{#1}}\kern0.25em%
    \ifcase #2\or%
        January\or February\or March\or%
        April\or May\or June\or%
        July\or August\or September\or%
        October\or November\or December%
    \fi\ %
    \number #3%
}

\usepackage[colorlinks=true,citecolor=blue]{hyperref}  % Needs to go last

\newcommand{\uo}[1]{u_{\mathrm{o}({#1})}}
\newcommand{\ut}[1]{u_{\mathrm{t}({#1})}}
\newcommand{\SU}{\mathrm{SU}}
\newcommand{\curlyL}{{\mathscr L}}
\newcommand{\curlyP}{{\mathcal P}}
\newcommand{\gothicL}{{\mathfrak L}}
\renewcommand{\id}{\mathrm{I}_2}
\newcommand{\hypergeom}[3]{\vphantom{F}_1F_1(#1\,; #2\,; #3)}

\newcommand{\bulletpoint}{\textbullet}

\newcommand{\UI}{\mathrm{I}}
\newcommand{\CG}{\mathcal{G}}
\newcommand{\CV}{\mathcal{V}}
\newcommand{\CB}{\mathcal{B}}
\newcommand{\ui}{\mathrm{i}}
\newcommand{\ue}{\mathrm{e}}
\renewcommand{\vecalpha}{\vecgamma}
\newcommand{\rmo}{\mathrm{o}}
\newcommand{\rmt}{\mathrm{t}}
\newcommand{\rmT}{\mathrm{T}}

\begin{document}
\title{Intermediate statistics for a system with symplectic
symmetry: the Dirac rose graph}
\author{J.M.~Harrison}
\affil{Department of Mathematics, Baylor University, One Bear Place,
Waco, Texas 76798, U.S.A.}
\author{B.~Winn}
\affil{Department of Mathematical Sciences,
Loughborough University, Loughborough,
LE11 3TU, U.K.}
\date{\printdate{3}{7}{2012}}
\maketitle
\begin{abstract}
  We study the spectral statistics of the Dirac operator on a
  rose-shaped graph---a graph with a single vertex and all bonds
  connected at both ends to the vertex. We formulate a secular
  equation that generically determines the eigenvalues of the Dirac
  rose graph, which is seen to generalise the secular equation for a
  star graph with Neumann boundary conditions. We derive
  approximations to the spectral pair correlation function at large
  and small values of spectral spacings, in the limit as the number of
  bonds approaches infinity, and compare these predictions with
  results of numerical calculations.
Our results represent the first example of intermediate statistics
from the symplectic symmetry class.
\end{abstract}

\thispagestyle{empty}

\section{Introduction}
For classically
chaotic quantum systems with an anti-unitary symmetry (for
example, time reversibility), the spectral statistics are conjectured
to be of random-matrix type from the Gau\ss ian Orthogonal Ensemble
\cite{boh:coc, cas:otc}. If the motion is generated by a Hamiltonian
of a half-integer spin particle, then instead the statistics of the
Gau\ss ian Symplectic Ensemble are expected \cite{sch:kda, cau:elr}.
This is a strong form of universality, where a single exact expression for
each spectral statistic is shared by all members of the universality class.

Between the dynamical extremes of full chaos, and complete integrability,
one can find quantum systems with a variety of \emph{intermediate}
spectral statistics.
Systems with intermediate statistics exhibit properties such as linear
level repulsion and an exponential decay in the probability for large
spacings.  However, as was pointed out in  \cite{gor:com} they do not share
precisely the same spectral statistics.

Prominent examples that have been investigated in recent years include
\cite{shk:sos} the Anderson model at the metal-insulator transition point,
\cite{dat:rbi, rah:ssot, bog:poc} Aharonov-Bohm integrable billiards,
\cite{seb:wcis, seb:wciq, alb:wci} rectangular billiards, or other
integrable dynamical system, perturbed by the addition of a point
singularity (this example has often  been called the \textit{\v{S}eba billiard})
and \cite{bog:moi, gre:sdf, par:nus} several other examples such as
polygonal billiards with rational angles.
(In \cite{tud:ssr} an alternative construction of the
operator corresponding to the \v{S}eba billiard is given which
exhibits Poissonian spectral statistics.)

Some other situations in which intermediate spectral statistics appear
include \cite{gir:isi} the intermediate statistics of eigenphases of
quantum maps, and \cite{bog:srp} a one-dimensional model for
intermediate statistics formed of a gas of energy levels interacting
with a logarithmic potential.

Spectral statistics of \v{S}eba billiards (see also \cite{wea:ros, shi:lsd})
have been thoroughly investigated
in \cite{bog:sso, bog:ss, bog:nnd, bog:sco, rah:sso, rah:ppo}.
This model is particularly note-worthy since it constitutes an
example of a member of a
mini universality class of integrable systems perturbed by a
rank-one singularity. Belonging to this class are quantum Neumann star
graphs \cite{ber:tps, ber:sga}, \v{S}eba billiards
and quantum Neumann rose graphs (see section~\ref{sec:neumann_rose} below).

As a measure of the spectral correlations we will mainly
be using the pair correlation
function $R_2(x)$ \cite{boh:cma}. For a spectrum $\{ \lambda_n \}$, scaled so
that the mean spacing is $1$, we define $R_2(x)$ by
\begin{equation}
  \label{eq:R2_def}
  \lim_{N\to\infty} \frac1N \sum_{m=0}^N \sum_{n=0}^N g(\lambda_n-\lambda_m) =
g(0) + \int_{-\infty}^{\infty} g(x)R_2(x)\,\rmd x,
\end{equation}
where $g$ belongs to an appropriate class of test functions. The first
term on the right-hand side of \eqref{eq:R2_def} comes from the
diagonal terms $m=n$.  By considering $g$ to be approximately an indicator
function we see that $R_2(x)$ is a measure of pairs of levels
(regardless of ordering) that lie within a given distance of each
other. Note that unlike the nearest-neighbour spacing statistic that is
sometimes studied, $R_2(x)$ is not a probability density.

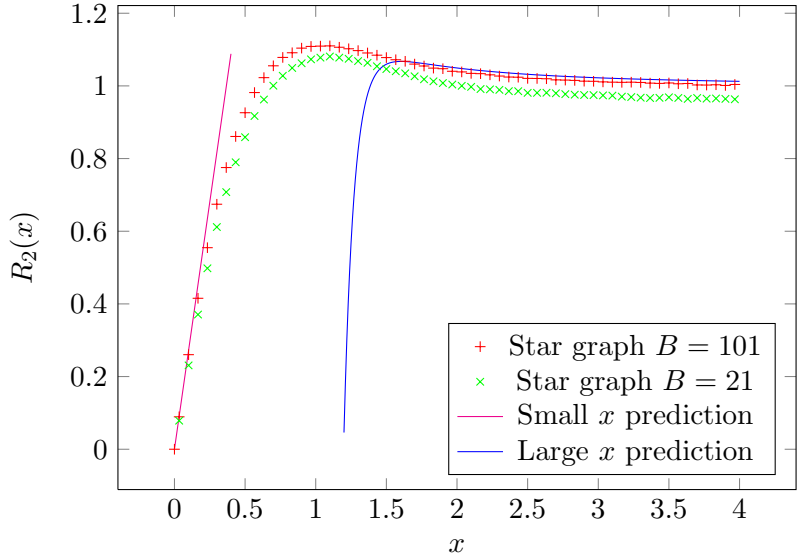
\begin{figure}[h]
\begin{center}
% \setlength{\unitlength}{6cm}
% \begin{picture}(1.333,1)
% \put(0,0){\includegraphics[angle=0,width=8cm,height=6cm]{PLOTS/star100-pcf-101-21-150K.pdf}}
% \end{picture}
  \begin{tikzpicture}
    \begin{axis}[
      xlabel=$x$,
      ylabel=$R_2(x)$]
      \addplot[color=red,
               mark=+,
               only marks] file {star100-pcf-101-150K.dat};
      \addlegendentry{Star graph $B=101$}
      \addplot[color=green,
               mark=x,
               only marks] file {star100-pcf-21-150K.dat};
      \addlegendentry{Star graph $B=21$}
      \addplot[color=magenta, no markers]
               coordinates {(0,0) (0.4,1.088279619)};
      \addlegendentry{Small $x$ prediction}
      \addplot[color=blue, no markers] file {starpaircorr-big.dat};
      \addlegendentry{Large $x$ prediction}
    \end{axis}
  \end{tikzpicture}
\caption{The pair correlation function numerically calculated for a $B=21$ bond
(green) and $B=101$ bond (red) star graph with Neumann boundary conditions.
Also plotted are the curves corresponding to the
large and small values of the parameter, \eqref{eq:R2_star_big} and
\eqref{eq:R2_star_small} respectively. The numerical plots are averaged over
100 realisations of the bond lengths, and 150\,000 eigenvalues were computed.}
\label{fig:uno}
\end{center}
\end{figure}

The spectral statistics that are shared by Neumann star graphs and \v{S}eba
billiards have been studied in some detail.
The pair correlation function for small $x$
was studied in \cite{bog:moi, bog:ss} and it was observed that it
behaves as
\begin{equation}
  \label{eq:R2_star_small}
  R_2(x) \approx\frac{\pi\sqrt3}2x \qquad\mbox{as $x\to 0$.}
\end{equation}
For the large $x$ asymptotics, a full series expansion has been derived
\cite{ber:tps, ber:sga}. Keeping terms up to $x^{-12}$ the expansion reads
\begin{equation}
  \label{eq:R2_star_big}
  R_2(x) = 1 + \frac{2}{\pi^2 x^2} + \frac{76}{\pi^4 x^4} -
\frac{1088}{\pi^6 x^6} + \frac{9280}{\pi^8 x^8} -
\frac{64000}{\pi^{10}x^{10}} + \Ord\!\left(\frac1{x^{12}}\right),
\qquad\mbox{as $x\to\infty$.}
\end{equation}
These behaviours are illustrated in figure \ref{fig:uno}, in which they are
compared to a numerical calculation of the pair-correlation function.
% star100-pcf-101-21-150K.pdf

Our motivation is to analyse the spectral statistics of a system in the
\v{S}eba class, with a symplectic symmetry. The most convenient model to
use for this investigation are quantum graphs, since it is known that
quantum star graphs with Neumann boundary conditions have statistics
in the \v{S}eba class, and Bolte and Harrison \cite{bol:ssf,bol:tsc}
successfully quantised
quantum graphs with the Dirac operator, and showed that generically the
statistics of the Gau\ss ian symplectic ensemble are found.

\begin{figure}
\centering
\begin{tikzpicture}[thick]
%
% A rose graph with 5 bonds:
%
\filldraw (0,0) circle (3pt);
\draw (0,0) .. controls (-2,2) and (0,3) .. (0,0);
\draw (0,0) .. controls (1,4) and (4,1) .. (0,0);
\draw (0,0) .. controls (3,0) and (2,-3) .. (0,0);
\draw (0,0) .. controls (1,-3) and (-1,-3) .. (0,0);
\draw (0,0) .. controls (-3,1) and (-2,-3) .. (0,0);
%
% A star graph with 6 bonds:
%
\filldraw (7,0) circle (3pt);
\foreach \x in {(7.2,2.5), (8.6,2), (8.3,-1), (5.5,-1), (5.3,0.5)} %  (7.2,-2)
 {
  \draw (7,0) -- \x;
  \filldraw \x circle (3pt);
 }
%
% Labels
%
\node at (-2.3,0) {(a)};
\node at (4.5,0) {(b)};
\end{tikzpicture}
\caption{(a) A rose graph with 5 bonds; (b) a star graph with 5 bonds.}
\label{fig:graphs}
\end{figure}
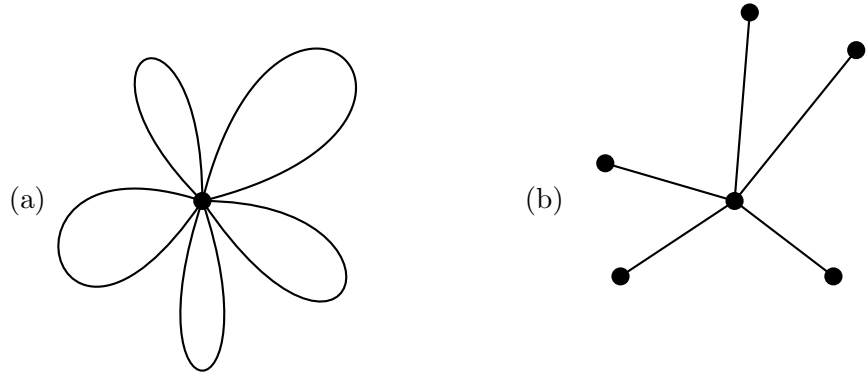

The only obstacle to this programme is that the construction of
\cite{bol:ssf} does not allow graphs with vertices of valency one, which
is most of the vertices of a star graph. For this reason we
re-attach the
loose ends of the star graph to the central vertex, to form a graph
which we call a \emph{rose graph} (see figure \ref{fig:graphs}).
We will see (section \ref{sec:neumann_rose}) that despite this change in the graph topology important features of the spectral analysis for star graphs survive the transformation.

The article is laid out in the following way.  In section
\ref{sec:quantisation} we introduce the general scheme used to
quantise the rose graph with the Dirac operator, in particular we
derive a simple secular equation whose roots provide the spectrum of
the graph with a similar structure to the well known secular equation
of the Neumann star graph.  Section \ref{sec:spectral statistics} presents the
derivation of the small and large parameter asymptotics of the pair
correlation function for the Dirac rose graph.  Section
\ref{sec:conclusions} draws together the results and compares them
with numerical calculations of the Dirac rose graph and a Neumann
star graph.

\section{Quantisation of rose graphs with the Dirac operator}
\label{sec:quantisation}

A graph $\CG$ consists of a set of vertices $\CV$ with pairs of
vertices connected by bonds, as in figure \ref{fig:graphs}.  Two
vertices $u,v\in \CV$ are adjacent $u\sim v$ if the pair $(u,v)$ is in
the set of bonds $\CB$, which may be associated with the set
$\{1,2,\ldots,B\}$.  We will also use $u\sim b$ and $v\sim b$ to show
that the bond $b$ is connected to $u$ and $v$.
%The number of vertices $u$ with $u\sim v$ is called the degree $d_v$ of $v$.
Each bond $b$ is associated with an interval $[0,L_b]$ where $L_b$ is
the length of $b$.  $\curlyL=2\sum_{b=1}^B L_b$ will denote twice the
total length of the graph; a natural measure of the volume of the
graph as each bond can be traversed in two directions.  We will assume
each $L_b$ lies in an interval $[1-(2B)^{-1},1+(2B)^{-1}]$ and that
the set of bond lengths are rationally independent.  On $[0,L_b]$ we
define a coordinate $x_b$ such that $x_b=0$ at the vertex $\rmo (b)$
and $x_b=L_b$ at the vertex $\rmt (b)$; $\rmo (b)$ and $\rmt (b)$ are
called the origin and terminus of $b$ respectively.  We will use $x$
to denote a general coordinate on the graph when the particular bond
is not significant.  The number of bonds $b$ with $\rmo (b)=v$ or
$\rmt (b)=v$ is $d_v$ the degree of the vertex $v$.  Note that a bond
$b$ with $\rmo (b)=\rmt (b)=v$, as seen in the rose graph, counts
twice when determining the degree of $v$.  It is sometimes convenient
to have notation for the reverse of a bond; $\overline{b}$ is a bond
with $\rmo (\overline{b})=t(b)$ and $\rmt (\overline{b})= \rmo(b)$ ($L_{\overline{b}}=L_b$).

On an interval $[0,L_b]$ the time independent Dirac equation reads,
\begin{equation}
  \label{eq:dirac_eq}
  -\rmi \alpha \frac{\rmd\vecpsi_b}{\rmd x_b} + m\beta \vecpsi_b = E \vecpsi_b
\ ,
\end{equation}
where $\alpha$ and $\beta$ satisfy the relations
$\alpha^2=\beta^2=\UI$ and $\alpha\beta+\beta\alpha=0$ which define
the Dirac algebra in one dimension.  The dimension of $\alpha$ and
$\beta$ depends on the interpretation of the Dirac equation in one
dimension.  If one looks for the simplest faithful irreducible
representation of the Dirac algebra $\alpha$ and $\beta$ will be
$2\times 2$ matrices acting on two component spinors.  On the other
hand, if one regards the equation in one dimension as the
restriction to a wire of a Dirac equation in three dimensions it is
natural to expect four component spinors.  It turns out that these two
different approaches lead to the same scattering problem
\cite{bol:ssf}.  In order to impose time-reversal symmetry with two
component spinors it is necessary to work with pairs of bonds
connecting vertices, effectively reintroducing four components to the
wave function for each adjacent pair of vertices.  To simplify the
current presentation we will only consider four component spinors from
the outset and fix
\begin{equation}
  \label{eq:A_and_B}
  \alpha = \left(\begin{array}{rrrr}
      0 & 0 & 0 & -\rmi \\
      0 & 0 & \rmi & 0 \\
      0 & -\rmi & 0 & 0 \\
      \rmi & 0 & 0 & 0
    \end{array}\right)
\qquad\text{and}\qquad
  \beta = \left(\begin{array}{rrrr}
      1 & 0 & 0 & 0 \\
      0 & 1 & 0 & 0 \\
      0 & 0 & -1 & 0 \\
      0 & 0 & 0 & -1
    \end{array}\right) \ .
\end{equation}

Given the operator on the collection of intervals we must now specify
a domain for which it is self-adjoint.  To
do this we fix appropriate matching conditions between the spinors on
the set of intervals where they meet at the vertices of the $\CG$.
Such matching conditions were classified in \cite{bol:ssf}.  In
particular there is a natural generalisation of matching conditions
that define a self-adjoint Laplace operator on the graph to conditions
for a Dirac operator.

%\marginpar{Include description of time-reversibility? \dbend}

For comparison the time-independent free Schr\"odinger equation
on an interval reads,
\begin{equation}
  \label{eq:chrodinger_eq}
  - \frac{\rmd^2 \psi_b}{\rmd x_b^2} = E \psi_b \ .
\end{equation}
Matching conditions between the functions on the individual intervals
where they meet at a vertex $v$ are expressed as $d_v$ linear
relations amongst the values of the functions $\psi_b$ and their
derivatives $\psi'_b$ evaluated at the end of each interval connected
to $v$.  For example, a common choice of matching conditions for the
Laplace operator are Neumann conditions; $\psi$ is continuous at each
vertex $v$;
\begin{align}\label{eq:Neummann continuity Laplace}
    \psi_b(0)&=\gamma \textrm{ for all }b \textrm{ with } \rmo(b)=v \ , \\
    \psi_b(L_b)&=\gamma \textrm{ for all }b \textrm{ with } \rmt(b)=v \ ,
\end{align}
(Note that $\gamma$ is not a fixed constant but rather a placeholder
for the value of the wave function at the vertex $v$ whatever that
happens to be) and the outgoing derivatives at $v$ sum to zero,
\begin{equation}\label{eq:Neumann conditions}
 \sum_{b\,|\,\rmo(b)= v} \psi_b'(0) -\sum_{b\,|\,\rmt(b)=v} \psi'_b(L_b)=0 \ .
\end{equation}
Fixing such matching conditions at each vertex of the graph ensures
that the Laplace operator is self-adjoint.  A full description of all
matching conditions for which the operator is self-adjoint has been
obtained in various forms \cite{har:hsg,kos:krf,kuc:qgI}, however for
the current discussion an understanding of Neumann conditions is sufficient.

For the Laplace operator it is clear that a wave function restricted to
bond $b$ can be written as a linear superposition of two plane waves,
\begin{equation}\label{eq:schrod wavefunction}
    \psi_b(x_b) = Q_b \ue^{\ui k x_b} + R_b \ue^{-\ui k x_b}  \ ,
\end{equation}
where $k^2=E$.  For simplicity we can consider the centre of the star
$v$ where $v=\rmo(b)$ for every $b\sim v$.  Then
$\vec{Q}=(Q_1,\dots,Q_B)$ and $\vec{R}$, defined similarly, are
vectors of the coefficients of incoming and outgoing plane wave
solutions at $v$.  Using the matching conditions these can be related
by a unitary matrix $\sigma^{(v)}(k)$ the vertex scattering matrix,
\begin{equation}\label{eq:vertex scattering matrix}
    \vec{R}=\sigma^{(v)}(k) \vec{Q} \ .
\end{equation}
With Neumann conditions at $v$ the vertex scattering matrix takes a
particularly simple form, see e.g.\ \cite{kot:pot},
\begin{equation}\label{eq:Neumann vertex scattering matrix}
    \sigma^{(v)}_{bc}=\left\{ \begin{array}{ccl}
    \frac{2}{d_v}&& \textrm{if }b\neq c \\
    \frac{2}{d_v}-1 && \textrm{if } b=c \\
    \end{array}
    \right. \ .
\end{equation}
Components of $\sigma^{(v)}$ are scattering amplitudes relating an
incoming plane wave on bond $c$ to an outgoing plane wave on bond $b$.
Time-reversal symmetry implies that $\sigma^{(v)}=(\sigma^{(v)})^{\mathrm{T}}$.

To quantise a graph with two component spinors one may treat the
spinor matching conditions by analogy with those of the Laplace
operator.  The first component of the spinors evaluated at the ends of
the bonds takes the place of the value of the scalar wave function of
the Laplace operator while the derivative of the wave function
evaluated at the ends of the intervals is replaced by the value of the
second component of the spinor at the bond ends. %, see \cite{bol:ssf}.
With the four component spinors, necessary to incorporate
time-reversal symmetry, the story is much the same.  However, now a
vector of the first and second components of the spinor is treated by
analogy with values of the wave function and the third and fourth
components replace values of the derivative. In the following
we summarise the relevant results from \cite{bol:ssf} concerning the
quantisation of a graph with the Dirac operator.

If $\vecpsi=(\psi_1, \ldots, \psi_4)^{\mathrm T}$ is a 4-component
spinor satisfying the Dirac equation (\ref{eq:dirac_eq})
it can be expressed as a linear combination of four plane waves two traveling
in each direction along the bond.  For $E>m$, $\vecpsi_b$ is of the form
\begin{multline}
  \label{eq:dirac_sol}
  \vecpsi_b = Q_b\left(\begin{array}{c} -\rmi \sqrt{E+m} \\ 0 \\ 0 \\
 \sqrt{E-m} \end{array}\right)\rme^{\rmi k x_b} +
  P_b\left(\begin{array}{c} 0 \\ \sqrt{E+m} \\ -\rmi\sqrt{E-m} \\ 0
      \end{array}\right)\rme^{\rmi k x_b} \\+
  R_b\left(\begin{array}{c} 0 \\ \sqrt{E+m} \\ \rmi\sqrt{E-m} \\ 0
      \end{array}\right)\rme^{-\rmi k x_b} +
  S_b\left(\begin{array}{c} \rmi\sqrt{E+m} \\ 0 \\ 0 \\ \sqrt{E-m}
      \end{array}\right)\rme^{-\rmi k x_b},
\end{multline}
where $k^2=E^2-m^2$, and $P_b, Q_b, R_b, S_b$ are constants of integration.
It is convenient to define
\begin{equation}
  \label{eq:Q_and_R}
  \vec{Q}_b=\left(\begin{array}{r} -\rmi Q_b \\ P_b \end{array}\right) \qquad
\text{and}\qquad
  \vec{R}_b=\left(\begin{array}{r} \rmi S_b \\ R_b \end{array}\right) \ .
\end{equation}
Then vectors of coefficients of the incoming and outgoing waves at the vertex
of a star can be written as,
\begin{equation}\label{eq:Dirac vector of incoming coeffs}
    \vec{Q}=(\vec{Q}_1^\rmT,\dots,\vec{Q}_B^\rmT)^\rmT \qquad \textrm{and}
\qquad
    \vec{R}=(\vec{R}_1^\rmT,\dots,\vec{R}_B^\rmT)^\rmT \ .
\end{equation}
As in the scalar case, vectors of the incoming and outgoing coefficients can
be related via a vertex scattering matrix;
\begin{equation}\label{eq:spinor vertex scattering matrix}
    \vec{R}=\sigma^{(v)}(k) \vec{Q} \ .
\end{equation}
Time-reversal symmetry requires that
\begin{equation}\label{eq:3tr2}
(\sigma^{(v)})^{\rmT}=  J^{-1} \sigma^{(v)} J \ ,
\end{equation}
where $J$ is a block diagonal matrix
\begin{equation}\label{eq:defn J}
J= \left( \begin{array}{ccc}
\left( \begin{array}{cc}
0 & 1 \\
-1 & 0 \\
\end{array} \right) && \\
& \ddots & \\
&&
\left( \begin{array}{cc}
0 & 1 \\
-1 & 0 \\
\end{array} \right) \\
\end{array} \right) \ .
\end{equation}
If the vertex scattering matrix is also divided into $2\times 2$
blocks $(\sigma^{(v)})^{bc}$, which relate pairs of incoming spinors
on bond $b$ to outgoing pairs on bond $c$, the symmetry condition
(\ref{eq:3tr2}) reads
\begin{equation}
(\sigma^{(v)})^{cb}= \det\left( (\sigma^{(v)})^{bc} \right)  \,
\left( (\sigma^{(v)})^{bc} \right) ^{-1} \ .
\end{equation}
We can factor a block of the vertex scattering matrix so
\begin{equation}\label{eq:factor scattering}
    (\sigma^{(v)})^{bc}= X_{bc} u_{bc} \ ,
\end{equation}
where $u_{bc} \in \SU (2)$, the symmetry condition is then $X_{cb}=X_{bc}$
and $u_{cb}=u_{bc}^{-1}$.  In this notation spinor scattering on the
graph is defined by a scalar scattering amplitudes $X_{bc}$ which
define the transition probability from bond $b$ to $c$, just as in the
case of scalar wave functions, and an
additional rotation between the spinors when passing through the
vertex, $u_{bc}$.  In general these transition amplitudes and spinor
rotations are obtained from the matching conditions and will also
satisfy the constraint that the whole scattering matrix $\sigma^{(v)}$
is unitary.  One straightforward way to satisfy these symmetry
relations also establishes a connection with the scattering matrices
of the Schr\"odinger operator. A vertex scattering matrix
of the form
\begin{equation}\label{eq:3gen}
\sigma^{(v)}= U^{(v)}
\left\{  X^{(v)} \otimes \UI_2
\right\}
\left( U^{(v)} \right)^{-1} \ ,
\end{equation}
where $X^{(v)}$ is a symmetric unitary $d_v\times d_v$ matrix and
$U^{(v)}=\textrm{diag}\{ u_1,\dots, u_{d_v} \}$ with $u_b\in \SU (2)$
satisfies the symmetry condition (\ref{eq:3tr2}).
The $d_v$ matrices $u_b$ can be thought of as defining a spinor
rotation when leaving the vertex on bond $b$ or the inverse rotation
when entering the vertex from that bond.  A symmetric unitary $d_v
\times d_v$ matrix of transition amplitudes $X^{(v)}$ is precisely the form of
vertex scattering matrix obtained for the Laplace operator with
time-reversal symmetry (\ref{eq:Neumann vertex scattering matrix}).
In fact, given matching conditions that make
the Laplace operator self-adjoint with a $d_v\times d_v$ vertex
scattering matrix $X^{(v)}$ we can define analogous matching
conditions which will make a Dirac operator self-adjoint for any given
set of spinor rotations $\{ u_1,\dots, u_{d_v} \}$.  For instance
matching conditions analogous to the Neumann conditions defined for
the Laplace operator (which with a slight abuse of terminology we will
also call Neumann conditions) are
\begin{align}
  \label{eq:bc_continuity_jon1}
\uo{b} \vec{v}_b(0) &= \vecalpha \textrm{ for all }b \textrm{ with }\rmo(b)=v \\
\ut{b} \vec{v}_b(L_b) &=\vecalpha \textrm{ for all }b \textrm{ with }\rmt(b)=v
  \label{eq:bc_continuity_jon2}
\end{align}
where on each bond $b$ we have such a solution $\vecpsi_b = (\psi_{b1},\ldots
\psi_{b4})^{\mathrm T}$ to \eqref{eq:dirac_eq} and we define
\begin{equation}
  \label{eq:28}
  \vec{v}_b(x) = \left( \begin{array}{c} \psi_{b1}(x) \\
   \psi_{b2}(x) \end{array}\right)
\qquad\text\qquad
  \vec{w}_b(x) = \left( \begin{array}{c} -\psi_{b4}(x) \\
      \psi_{b3}(x) \end{array}\right) \ .
\end{equation}
In (\ref{eq:bc_continuity_jon1})--(\ref{eq:bc_continuity_jon2})
$\vecalpha$ is again a placeholder
for a common value of the spinor at the vertex rather than a fixed
constant vector.  In addition the Neumann condition on the derivatives
(\ref{eq:Neumann conditions}) becomes,
\begin{equation}
  \label{eq:bc_kirchhoff first}
 \sum_{b\,|\,\rmo(b)=v} \uo{b} \vec{w}_b(0) -
\sum_{b\,|\,\rmt(b)=v} \ut{b}\vec{w}_b(L_b) = 0 \ .
\end{equation}
Together applying these conditions at all vertices of a graph defines a
self-adjoint realization of the Dirac operator \cite[section 5]{bol:ssf}.

To each bond $b$ of a graph we now see that there correspond a pair of
$2\times 2$ unitary matrices $\uo{b}$ and $\ut{b}$ which rotate the
spinor when entering the vertices at the origin and terminus of $b$
respectively.  At a vertex, during the transition from a bond $b_j$ to
a bond $b_i$ the spinor is rotated according to the matrix
$u_{b_ib_j}\in\SU(2)$ where
\begin{equation}
  \label{eq:v_matrix}
  u_{b_ib_j} \coloneq \uo{b_i}\left( \ut{b_j} \right)^{-1},
\end{equation}
%(see \cite{bol:ssf} for details).
It will be convenient to associate
spin matrices to \emph{bonds} rather than vertices, so for this reason
we define
\begin{equation}
  \label{eq:w_matrix}
  \tilde{w}_b \coloneq \uo{b}\left(\ut{b}\right)^{-1}
\end{equation}
and
\begin{equation}
  w_b \coloneq \left( \ut{b} \right)^{-1} \uo{b}.
\end{equation}
We note that if $\bar{b}$ is the reversal of the bond $b$,
\begin{align}
  w_{\bar{b}} &= \left( \ut{\bar{b}} \right)^{-1} \uo{\bar{b}} \nonumber \\
  &= \left( \uo{b} \right)^{-1} \ut{b} \nonumber \\
  &= w_b^{-1}. \label{eq:w_inverse}
\end{align}
If the $2B$ matrices $\{ \uo{b}, \ut{b} \}_{b\in{\mathcal B}}$ are
random and independently distributed, then the matrices
$\{ w_b\}_{b\in{\mathcal B}}$ are independently distributed. If the
matrices $\{ \uo{b}, \ut{b} \}_{b\in{\mathcal B}}$ are independently
distributed with Haar measure on $\SU(2)$, then the $w_b$ are
Haar-distributed too, and the angles $\{ \theta_b \}_{b\in{\mathcal B}}$
defined on $[0,\pi]$ by
\begin{equation}
  \label{eq:def_theta}
  \tr\tilde{w}_b = \tr w_b \eqcolon 2\cos\theta_b,
\end{equation}
are identically independently distributed with a sine-squared distribution:
\begin{equation}
  \label{eq:sine_squared}
  \Prob\left(\theta_b < x \right) = \frac2\pi \int_0^x \sin^2\phi\,\rmd\phi,
\qquad 0\leq x \leq \pi.
\end{equation}
The traces themselves have a semi-circle distribution:
\begin{equation}
  \label{eq:semi_circle}
\Prob\left(\tr w_b < x \right) = \frac1{2\pi} \int_{-2}^x \sqrt{4-t^2}\,\rmd t,
\qquad -2\leq x \leq 2.
\end{equation}

Bolte and Harrison proved a trace formula for the eigenvalues of the
Dirac operator on an arbitrary graph in terms of periodic orbits
\cite{bol:ssf}. In order to state the trace formula, we need to define
a few quantities.  A periodic orbit $p$ is a closed path on the graph,
modulo cyclic shifts. So $p$ can be written as a sequence of connected
bonds $p=(b_1,b_2,\dots,b_n)$ where $\rmt(b_j)=\rmo(b_{j+1})$ and
$\rmt(b_n)=\rmo(b_0)$. We denote by the symbol $\curlyP$ the family of
all periodic orbits on the graph.  It is possible that an orbit $p\in
\curlyP$ can be written as a repetition of a shorter periodic orbit,
in which case $r_p$, the repetition number of $p$, is the maximal
number of repetitions of a shorter orbit contained in $p$.
$A_p$ will denote the product of all scattering amplitudes accumulated
along the orbit.  For the free Schr\"odinger operator this is simply
be products of the elements of the vertex scattering matrices;
\begin{equation}\label{eq:stability amplitude}
    A_p=\sigma^{(\rmt(b_1))}_{b_2b_1} \sigma^{(\rmt(b_2))}_{b_3b_2} \dots
\sigma^{(\rmt(b_n))}_{b_1b_n} \ .
\end{equation}
While for the Dirac operator we replace $\sigma^{(\rmt(b_j))}_{b_{j+1}b_j}$ with $X^{(\rmt(b_j))}_{b_{j+1}b_j}$.
We will denote by $d_p$ the matrix
product of all spin matrices accumulated on the orbit:
\begin{equation}
  \label{eq:11}
  d_p \coloneq u_{b_nb_{n-1}} u_{b_{n-1}b_{n-2}} \cdots u_{b_2b_1} u_{b_1b_n}
\in \SU(2).
\end{equation}
Because of the cyclic invariance of the trace, we have
\begin{equation}
  \label{eq:12}
  \tr d_p = \tr( w_{b_n} w_{b_{n-1}} \cdots w_{b_2} w_{b_1} ).
\end{equation}
$\ell_p$ denotes the length of the orbit (i.e.\ the sum of lengths of
bonds over which the orbit passes).

Let $h$ be a test
function whose Fourier transform $\hat{h}(\ell) = \int_{-\infty}^\infty
h(k)\rme^{-2\pi \rmi k \ell}\,\rmd k$ is smooth and compactly
supported. Then, in terms of quantities defined above, the
trace formula reads:
\begin{equation}
  \label{eq:trace_formula}
  \sum_{n=1}^\infty h(k_n) = \frac{\curlyL}{\pi}H(0) + \frac2\pi
\sum_{p\in\curlyP} \frac{\ell_p}{r_p}\left( \frac{\tr d_p}2 \right)
A_p H(\ell_p),
\end{equation}
where
\begin{equation}
  H(\ell) \coloneq \int_{-\infty}^\infty h(k)\cos k\ell \,\rmd k.
\end{equation}
In \eqref{eq:trace_formula} recall that the quantity $\curlyL$ is
twice the total length of the graph.

\subsection{Dirac rose graphs} \label{sec:zwei_punkt_eins}

On the rose graph there are $B$ intervals with both ends of each
interval connected to the single central vertex, figure
\ref{fig:graphs}(b).  Eigenspinors on the bonds have the form shown in
equation (\ref{eq:dirac_sol}).  The Neumann matching conditions at the
central vertex, which will determine the constants of integration,
are,
\begin{equation}
  \label{eq:bc_continuity}
  \uo{b} \vec{v}_b(0) = \ut{b} \vec{v}_b(L_b) = \vecalpha\qquad
\text{for all $b$,}
\end{equation}
which imposes continuity of the first pair of components of the spinor, and
\begin{equation}
  \label{eq:bc_kirchhoff}
  \sum_{b=1}^B \uo{b} \vec{w}_b(0) = \sum_{b=1}^B \ut{b}\vec{w}_b(L_b).
\end{equation}
% which is the Kirchhoff condition. Together these conditions define a
% self-adjoint extension of the Dirac operator \cite[section 5]{bol:ssf}.

% It is convenient to define
% \begin{equation}
%   \label{eq:Q_and_R}
%   \vec{Q}_b=\left(\begin{array}{r} -\rmi Q_b \\ P_b \end{array}\right) \qquad
% \text{and}\qquad
%   \vec{R}_b=\left(\begin{array}{r} \rmi S_b \\ R_b \end{array}\right).
% \end{equation}

Using the solution \eqref{eq:dirac_sol} and \eqref{eq:Q_and_R},
condition \eqref{eq:bc_continuity} reads
\begin{equation}
  \label{eq:31}
  \uo{b}(\vec{Q}_b+\vec{R}_b) = \ut{b}(\vec{Q}_b\rme^{\rmi k L_b} + \vec{R}_b
\rme^{-\rmi k L_b}) = \frac{\vecalpha}{\sqrt{E+m}},
\end{equation}
which gives
\begin{equation}
  \label{eq:32}
  \left( \uo{b} - \ut{b}\rme^{\rmi k L_b}\right) \vec{Q}_b = -\left(
 \uo{b} - \ut{b}\rme^{-\rmi k L_b}\right)\vec{R}_b.
\end{equation}
From \eqref{eq:31},
\begin{equation}
  \label{eq:33}
 \frac{\uo{b}^{-1}\vecalpha}{\sqrt{E+m}} =  \vec{Q}_b+\vec{R}_b,
\end{equation}
and so, eliminating $\vec{R}_b$ using \eqref{eq:32} we get
% \begin{align}
%    \frac1{\sqrt{E+m}} (\uo{b}-\ut{b}\rme^{-\rmi k L_b}){\uo{b}^{-1}\vecalpha}
% &= (\uo{b} - \ut{b}\rme^{-\rmi k L_b} - \uo{b} + \ut{b}\rme^{\rmi k L_b})
% \vec{Q}_b \nonumber \\
% &= 2\rmi\sin kL_b \ut{b} \vec{Q}_b,
%   \label{eq:34}
% \end{align}
% so that
\begin{equation}
  \label{eq:vec_Q}
  \vec{Q}_b\sin kL_b = \frac1{2\rmi \sqrt{E+m}}\ut{b}^{-1} (\uo{b}-\ut{b}
\rme^{-\rmi k L_b}) \uo{b}^{-1}\vecalpha.
\end{equation}

Let us now, for simplicity, consider the case that $L_b k/\pi\not\in\Z$ for
every bond $b$.
Applying the boundary condition \eqref{eq:bc_kirchhoff} gives
\begin{equation}
  \label{eq:36}
  \sum_{b=1}^B -\rmi \uo{b}(\vec{Q}_b-\vec{R}_b) = \sum_{b=1}^B -\rmi \ut{b}
  (\vec{Q}_b \rme^{\rmi k L_b} - \vec{R}_b \rme^{-\rmi k L_b}),
\end{equation}
which simplifies to
\begin{equation}
  \label{eq:35}
  \sum_{b=1}^B 2\left( \uo{b}-\rme^{\rmi k L_b}\ut{b}\right)\vec{Q}_b = \vec{0}
\end{equation}
upon inserting \eqref{eq:32}. Substituting \eqref{eq:vec_Q}, we get
\begin{align}
\frac1{\rmi\sqrt{E+m}}\sum_{b=1}^B \frac1{\sin kL_b}\left( (\uo{b} -
 \rme^{\rmi k L_b}\ut{b}) \ut{b}^{-1} ( \uo{b} - \ut{b}\rme^{-\rmi k L_b})
\uo{b}^{-1} \right)\vecalpha &= \vec{0} \nonumber \\
\Rightarrow\quad \sum_{b=1}^B \frac1{\sin kL_b}\left(
  \uo{b}\ut{b}^{-1} - \rme^{\rmi k L_b}\id \right)\left( \id -
  \ut{b}\uo{b}^{-1}\rme^{-\rmi k L_b}\right) \vecalpha &= \vec{0} \nonumber \\
\Rightarrow\quad \sum_{b=1}^B \frac1{\sin kL_b} \left( \tilde{w}_b +
  \tilde{w}_b^{-1} - 2\cos kL_b\id \right) \vecalpha &=\vec{0}.
  \label{eq:37}
\end{align}
Since $\tilde{w}_b + \tilde{w}_b^{-1} = \tr{\tilde{w}_b}\id =
2\cos\theta_b\id$, the condition for $k$ to be an eigenvalue becomes
\begin{equation}
  \label{eq:eigenvalue_condition}
 \sum_{b=1}^B \frac{\cos\theta_b - \cos kL_b}{\sin kL_b} = 0.
\end{equation}
Equation \eqref{eq:eigenvalue_condition} is the secular equation for
a Dirac rose graph.

Let us now consider what happens if $L_bk/\pi = n\in\Z$ for some bond $b$. In
that case, \eqref{eq:vec_Q} may be re-written as
\begin{align}
  \label{eq:38}
  \vec{0} = \vec{Q}_b\sin kL_b
 &= \frac{1}{2\rmi\sqrt{E+m}} \ut{b}^{-1} (\uo{b} - (-1)^n \ut{b})
\uo{b}^{-1}\vecalpha \nonumber \\
&=\frac{1}{2\rmi\sqrt{E+m}} \ut{b}^{-1} (\tilde{w}_b - (-1)^n\id)
  \ut{b}\uo{b}^{-1}\vecalpha.
\end{align}
Thus, if $\tilde{w}_b \neq (-1)^m \id$, we find that
$\vecalpha=\vec{0}$, and consequently $\vec{Q}_c=\vec{R}_c=\vec{0}$
for all other $c\neq b$, since the rational independence of bond
lengths means that $L_c k/\pi\not\in\Z$. From the boundary condition
\eqref{eq:bc_kirchhoff}, we then would have
$\vec{Q}_b=\vec{R}_b=\vec{0}$, and there are consequently no
non-trivial solutions to \eqref{eq:dirac_eq}.

We therefore conclude that if all $\tilde{w}_b\neq\pm\id$, then the
eigenvalues $k_n$, $n\in\Z$ are given by the condition
\eqref{eq:eigenvalue_condition}, which is the central result of this
paper.
We also remark that if the matrices $\{ \uo{b}, \ut{b}\}_{b\in\mathcal{B}}$ are
chosen randomly independently with Haar measure, then almost-surely,
$\tilde{w}_b\neq\pm\id$.

Kramer's degeneracy is evident in equation \eqref{eq:37}, since for any
value of $k$ satisfying \eqref{eq:eigenvalue_condition}, there is a
$2$-dimensional space of solutions for $\vecalpha$. This leads to two
linearly independent eigensolutions, so each eigenvalue occurs with
multiplicity $2$.

\subsection{Neumann rose graphs} \label{sec:neumann_rose}
We can use \eqref{eq:eigenvalue_condition} to derive the condition for $k$ to
be an eigenvalue of a rose graph quantised with the Laplace operator and
Neumann boundary conditions. The key observation is that if we choose the
$\SU(2)$ matrices to be
\begin{displaymath}
  \uo{b} = \ut{b} = \left( \begin{array}{cc} 1 & 0 \\ 0 & 1 \end{array}\right)
\qquad\text{for all bonds $b$,}
\end{displaymath}
and choose $m=0$, then the spectral problem given by
\eqref{eq:dirac_eq} and \eqref{eq:bc_kirchhoff} degenerates into a pair of
independent copies of the Neumann eigenproblem for the Laplace operator.

Since $\tilde{w}_b = \id$ for all $b\in\mathcal{B}$ there will be
eigenvalues of the form $k=2n\pi/L_b$ for $n\in\Z$. The rational
independence of bond lengths implies that $L_c k/2\pi\not\in\Z$ for any other
$c\neq b$, so following the analysis of section \ref{sec:zwei_punkt_eins} we
get (c.f.\ equation \eqref{eq:37}),
\begin{equation}
  \label{eq:cond_neumann_0}
   \sum_{\substack{c=1\\c\neq b}}^B \frac1{\sin kL_c} \left( 2\id
  - 2\cos kL_c\id \right) \vecalpha =\vec{0}.
\end{equation}
So $\vecalpha=\vec{0}$, and thence $\vec{Q}_c=\vec{R}_c=\vec{0}$ for
all $c\neq b$.

The remaining points of the spectrum come from the condition
\eqref{eq:eigenvalue_condition}, which in this situation is
expressible as
\begin{equation}
  \label{eq:cond_neumann}
   \sum_{b=1}^B \frac{1 - \cos kL_b}{\sin kL_b} =
 \sum_{b=1}^B \tan\left( \frac{kL_b}2 \right) = 0.
\end{equation}
The equality here between the left-hand sides comes from the observation
that
\begin{equation}
  \label{eq:tan_identity}
  \frac{1-\cos\varphi}{\sin\varphi} = \tan\left(\frac{\varphi}2\right),\qquad
\text{if }\frac{\varphi}{2\pi}\not\in\Z.
\end{equation}
What is particularly interesting is that condition \eqref{eq:cond_neumann}
is exactly the eigenvalue condition for a Neumann star graph \cite{ber:sga} with set of bond
lengths $\{ L_b/2 : b\in\mathcal{B}\}$.

The spectrum $\{k_n\}$ of the Laplace operator on rose graphs with
Neumann boundary conditions consists of points $k=k_n$ satisfying
\eqref{eq:cond_neumann}, together with points of the form $k=2m\pi/L_b$
with $b\in\mathcal{B}$ and $m\in \Z$, which interlace the solutions
to \eqref{eq:cond_neumann}.
Eigenfunctions corresponding to the latter class of eigenvalues are
supported on a single bond $b$.

\section{Spectral statistics}\label{sec:spectral statistics}
For a Dirac rose graph the eigenvalues are almost-surely the solutions
$k$ to the equation
\begin{equation}
  \label{eq:spec_det}
  Z(k) \coloneq Z(k;\vec{L},\vectheta)
       \coloneq \sum_{b=1}^B \frac{\cos\theta_b - \cos(kL_b)}{\sin(k L_b)} = 0.
\end{equation}
Define
\begin{equation}
  \label{eq:z}
  z(x,\theta) \coloneq \frac{\cos\theta - \cos x}{\sin x},
\end{equation}
so that
\begin{equation}
  Z(k) = \sum_{b=1}^B z(kL_b,\theta_b).
\end{equation}

Let us note an alternative form for the function $z(x,\theta)$.
From \cite{gra:tis} (equations 1.421.3 and 1.422.3) we have the pole
expansions:
\begin{equation}
  \label{eq:cot}
  \cot z = \frac1z +2 \sum_{k=1}^\infty \frac{z}{z^2-k^2\pi^2}
\end{equation}
and
\begin{equation}
  \label{eq:cosec}
  \cosec z = \frac1z + 2\sum_{k=1}^\infty \frac{(-1)^kz}{z^2-k^2\pi^2}.
\end{equation}
Since
\begin{equation}
  \label{eq:u}
 \frac{z}{z^2-k^2\pi^2} = \frac12\left( \frac1{z+k\pi} - \frac{k\pi}{1+k^2\pi^2}
+\frac1{z-k\pi} + \frac{k\pi}{1+k^2\pi^2}\right),
\end{equation}
we have
\begin{align}
  z(x,\theta) &= \cos\theta\cosec x - \cot x \nonumber\\
  &=\sum_{m=-\infty}^\infty ((-1)^m\cos\theta - 1)\left( \frac1{x+\pi m} -
\frac{m\pi}{1+m^2\pi^2}\right).
\label{eq:new_z}
\end{align}
(Because of the regularisation, the series in \eqref{eq:new_z} converges
absolutely.)

From this representation, we see that the structure of $Z(k)$ on the real
axis is a sequence of poles at the points $k=m\pi/L_b$ for $m\in\Z$ and
$b=1,\ldots,B$. The only way that poles can become closely-spaced is if
poles of $z(kL_b,\theta_b)$ become close for \emph{different} bonds $b$.
If the set of bond-lengths $\{ L_1,\ldots,L_B\}$ is linearly independent over
$\Q$, then the positions of the poles for different $b$ become independent
as $k$ becomes large.

By differentiating \eqref{eq:spec_det}, we get
\begin{equation}
  \label{eq:diff_Z}
  Z'(k) = \sum_{b=1}^B L_b\left(\frac{1-\cos(k L_b)\cos\theta_b}{\sin^2
(kL_b)}\right) \geq 0,
\end{equation}
so that $Z(k)$ is increasing between poles, and there is a unique zero
of $Z(k)$ between each consecutive pair of poles.

% These observations form the basis of the calculation of an approximation to
% the small-spacings statistics for the eigenvalues of the Dirac rose graph
% in the next section. \marginpar{Refer to Bogomolny \dbend}

These observations mean that spectral statistics of Dirac rose graphs
fall into the class considered by Bogomolny \textit{et al.} in
\cite{bog:ss}.  By averaging over the random phases
$\{\theta_b\}_{b\in\mathcal{B}}$, their analysis can be
used---with only slight modifications---to derive an expression for
the averaged pair-correlation function $R_2(x)$ (see equation (142),
\textit{loc.\ cit.}). In the following two sections we derive an
approximation to $R_2(x)$ for small $x$, by following a method
developed in \cite{bog:ss} and other places; and for large $x$, by a
different method based on the trace formula \eqref{eq:trace_formula}.

\mathversion{bold}
\subsection{Small $x$ behaviour of $R_2(x)$}
\mathversion{normal}
We shall use the method from \cite{bog:moi, bog:ss, gor:com} (see also
\cite{bar:otl}) that begins with the observation that statistics of small
spacings of zeros of a random meromorphic function, real with poles on the
real axis, are approximated
by statistics of the zeros of a function with three randomly-distributed
poles:
\begin{equation}
  \label{eq:three_poles}
  \frac{A_1}{k-e_1}+\frac{A_2}{k-e_2}+\frac{A_3}{k-e_3} = 0,
\end{equation}
where $A_1,\ldots,A_3$ are random amplitudes, and $e_1,\ldots,e_3$ are
random points. We can assume that the $e_1,\ldots,e_3$ are distributed
over a wide interval of the real axis, since only close spacings will
contribute to the approximation for statistics of small zeros
spacings.

From \eqref{eq:spec_det} and \eqref{eq:new_z}, the amplitudes $A_j$
are given by the quantities $(-1)^{m_j}\cos\theta_j-1$ where
$m_j\in\Z$ and $\theta_j$ are independent identically distributed
random angles according to \eqref{eq:sine_squared}. More explicitly,
$A_j$ is a linear transformation of $\cos\theta_j$, so it follows
from \eqref{eq:semi_circle} that the probability density of each $A_j$
is
\begin{equation} \label{eq:A_density}
  \frac{2}{\pi} \sqrt{y(2-y)}\I_{[0,2]}(y).
\end{equation}

We follow the method of \cite{bog:ss} which was used for a similar
situation in which all $A_j$ are equal to $1$. We begin by
re-arranging \eqref{eq:three_poles}, to find that the solutions are
given by the quadratic equation:
\begin{equation}
(A_1+A_2+A_3)k^2 - (A_1(e_2+e_3) + A_2(e_1+e_3) + A_3(e_1+e_2))k +
A_1e_2e_3 + A_2e_1e_3 + A_3e_1e_2 = 0.
\end{equation}
Denoting the solutions by $k_{1,2}$, we have
\begin{equation}
  \label{eq:2}
  k_{1,2}= \frac{A_1(e_2+e_3) + A_2(e_1+e_3) + A_3(e_1+e_2)\pm\sqrt{{\mathcal
D}}}{2(A_1+A_2+A_3)},
\end{equation}
where
\begin{equation}
  \label{eq:3}
  {\mathcal D} = (A_1(e_2+e_3) + A_2(e_1+e_3) + A_3(e_1+e_2))^2 - 4
(A_1e_2e_3 + A_2e_1e_3 + A_3e_1e_2)(A_1+A_2+A_3).
\end{equation}
If $L$ is any real number, then the translation $(e_1,e_2,e_3)\mapsto
(e_1+L,e_2+L,e_3+L)$ shifts $k_1$ and $k_2$ by an amount $L$. It is therefore
convenient to shift to a set of co\"ordinates in which
\begin{equation}
  \label{eq:4}
  A_1(e_2+e_3) + A_2(e_1+e_3) + A_3(e_1+e_2) = 0.
\end{equation}
Then \eqref{eq:3} becomes
\begin{align}
  {\mathcal D} &= - 4
(A_1e_2e_3 + A_2e_1e_3 + A_3e_1e_2)(A_1+A_2+A_3) \nonumber \\
  \label{eq:5}
&=4 (A_1+A_2+A_3)\left( \frac{A_2(A_2+A_3)}{A_1+A_2}e_1^2 +
\frac{A_1(A_1+A_3)}{A_1+A_2}e_2^2 + \frac{2A_1A_2}{A_1+A_2}e_1e_2\right).
\end{align}
To get the leading contribution to the pair-correlation function, we average
over the positions $e_1$ and $e_2$, and the random amplitudes $A_1,\ldots,A_3$:
\begin{equation}
  \label{eq:pair_corr_sm}
\E(R_2(x))\approx \frac12\E\bigg\{ \int_{-\infty}^\infty \!\int_{-\infty}^\infty
 \delta( x - \Delta k)\,\rmd e_1\rmd e_2\bigg\},
\end{equation}
where
\begin{equation}
  \label{eq:7}
  \Delta k  = k_2-k_1 = \frac{2\sqrt{{\mathcal D}}}{2(A_1+A_2+A_3)},
\end{equation}
and ${\mathcal D}$ is given by \eqref{eq:5}. To perform the integral in
\eqref{eq:pair_corr_sm}, we switch to a system of polar co\"ordinates. Let
\begin{equation}
  \label{eq:1}
  e_1 = \sqrt{\frac{A_1+A_2}{A_2(A_2+A_3)}}\rho\cos\varphi \qquad
\mbox{and}\qquad e_2 = \sqrt{\frac{A_1+A_2}{A_1(A_1+A_3)}}\rho\sin\varphi.
\end{equation}
Then we get
\begin{multline}
 \E( R_2(x))\approx \frac12\E\bigg\{ \int_0^{2\pi}\!\!\int_0^\infty
\frac{A_1+A_2}{\sqrt{A_1A_2(A_1+A_3)(A_2+A_3)}}\rho\times \\
\delta\bigg( x - \frac{2\rho}{\sqrt{A_1+A_2+A_3}}\bigg( 1 +
\frac{2\sqrt{A_1A_2}}{\sqrt{(A_1+A_3)(A_2+A_3)}}\cos\varphi\sin\varphi
\bigg)^{1/2}\bigg)\,\rmd\rho\rmd\varphi\bigg\},
\end{multline}
which becomes, after a further substitution,
\begin{multline}
  \label{eq:6}
 \E( R_2(x))\approx\frac12\E\bigg\{ \frac{A_1+A_2}{\sqrt{A_1A_2(A_1+A_3)
(A_2+A_3)}}
\frac{A_1+A_2+A_3}4 \times\\
\int_0^{2\pi} \bigg(  1 +
\frac{2\sqrt{A_1A_2}}{\sqrt{(A_1+A_3)(A_2+A_3)}}\cos\varphi\sin\varphi
\bigg)^{-1}\rmd\varphi\int_0^\infty \rho\delta(x-\rho)\,\rmd\rho\bigg\}.
\end{multline}
We use the known integral formula
\begin{equation}
  \label{eq:8}
  \int_0^{2\pi}\frac{\rmd\varphi}{1+a\cos\varphi\sin\varphi} = \frac{2\pi}
{\sqrt{1-a^2/4}},\qquad\mbox{for $|a|<2$,}
\end{equation}
to get
\begin{align}
\int_0^{2\pi} \bigg(  1 +
\frac{2\sqrt{A_1A_2}}{\sqrt{(A_1+A_3)(A_2+A_3)}}\cos\varphi\sin\varphi
\bigg)^{-1}\rmd\varphi
 & = 2\pi\left( 1 - \frac{A_1A_2}{(A_1+A_3)(A_2+A_3)} \right)^{-1/2} \nonumber
\\
 & = \frac{2\pi\sqrt{(A_1+A_3)(A_2+A_3)}}{A_3^{1/2}\sqrt{A_1+A_2+A_3}}.
  \label{eq:9}
\end{align}
Substituting into \eqref{eq:6} we arrive to
\begin{equation}
  \label{eq:R2_sm_unaveraged}
 \E( R_2(x))\approx \frac{\pi x}4\E\bigg\{ \frac{(A_1+A_2)\sqrt{A_1+A_2+A_3}}
{\sqrt{A_1A_2A_3}}\bigg\}.
\end{equation}
As we are assuming that the $A_j$ are identically distributed, we may
symmetrize \eqref{eq:R2_sm_unaveraged} and get
\begin{equation}
  \label{eq:R2_sm_averaged}
  \E(R_2(x)) \approx\frac{\pi x}6\E\bigg\{ \frac{(A_1+A_2+A_3)^{3/2}}
{\sqrt{A_1A_2A_3}}\bigg\}.
\end{equation}
This co\"\i ncides with equation (155) from \cite{bog:ss}, where it was
stated but not explicitly derived.

We have to evaluate \eqref{eq:R2_sm_averaged} in the case that $A_1,\ldots, A_3$
are independent and identically distributed with probability density
\eqref{eq:A_density}. To do that, it is helpful to observe that since
the $A_j$ are identically distributed, we have
\begin{multline}
  \label{eq:39}
  \E\bigg\{ \frac{(A_1+A_2+A_3)^{3/2}}
{\sqrt{A_1A_2A_3}}\bigg\} = 3\E\bigg\{ \frac{A_1^{3/2}}{A_2^{1/2}A_3^{1/2}}
\frac1{\sqrt{A_1+A_2+A_3}} \bigg\} \\ + 6\E\bigg\{ \frac{A_1^{1/2}A_2^{1/2}}
{A_3^{1/2}}\frac1{\sqrt{A_1+A_2+A_3}} \bigg\}.
\end{multline}
Also, since $A_j>0$ almost-surely, we also have the identity
\begin{equation}
  \label{eq:A_identity}
  \frac1{\sqrt{A_1+A_2+A_3}} = \frac1{\sqrt\pi}\int_{-\infty}^\infty
\rme^{-(A_1+A_2+A_3)x^2}\,\rmd x.
\end{equation}
By Fubini's theorem, we have, from \eqref{eq:39},
\begin{align}
  \label{eq:41}
  \E\bigg\{ \frac{(A_1+A_2+A_3)^{3/2}}
{\sqrt{A_1A_2A_3}}\bigg\} &= \frac3{\sqrt\pi}\int_{-\infty}^\infty
\E\left\{ A_1^{3/2}A_2^{-1/2}A_3^{-1/2}\rme^{-(A_1+A_2+A_3)x^2}\right\}\,\rmd x
\nonumber\\
  &\qquad\qquad +\frac6{\sqrt{\pi}}\int_{-\infty}^\infty
\E\left\{ A_1^{1/2}A_2^{1/2}A_3^{-1/2}\rme^{-(A_1+A_2+A_3)x^2}\right\}\,\rmd x
\nonumber \\
&=\frac3{\sqrt\pi} \int_{-\infty}^\infty I_{3/2}(x) I_{-1/2}(x)^2 +
2 I_{1/2}(x)^2 I_{-1/2}(x)\,\rmd x,
\end{align}
where
\begin{equation}
  \label{eq:I_integral}
  I_\nu(x)\coloneq \E\left\{ A_1^\nu \rme^{-A_1x^2}\right\} =
  \frac{2}\pi \int_0^2 y^{\nu+1/2}(2-y)^{1/2} \rme^{-yx^2}\,\rmd y.
\end{equation}
From \cite{gra:tis} formula 3.383.1, we find that
\begin{equation}
  \label{eq:I_value}
  I_\nu(x) = \frac{2^{\nu+2}}{\sqrt\pi} \frac{\Gamma(\nu+3/2)}{\Gamma(\nu+3)}
  \hypergeom{\nu+{\textstyle\frac32}}{\nu+3}{-2x^2}.
\end{equation}
The asymptotic behaviour of the confluent hypergeometric function allows us
to conclude that
\begin{equation}
  \label{eq:integrand_estimate}
   I_{3/2}(x) I_{-1/2}(x)^2 + 2 I_{1/2}(x)^2 I_{-1/2}(x) = \Ord\!\left(
\frac1{x^{10}}\right) \qquad\text{as $x\to\pm\infty$,}
\end{equation}
so that the integral in \eqref{eq:41} converges quickly, and may be
accurately evaluated using numerical integration techniques.
So doing, we arrive at the following result:
\begin{equation}
  \label{eq:R2_rose_small}
  \E(R_2(x)) \approx \frac{\pi c}{6}x \qquad\text{as $x\to0$,}
\end{equation}
where
\begin{equation}
  \label{eq:c_value}
  c \approx 6.781\ldots  % Exactly 6.7806799   --- see note below.
\end{equation}
The numerical value for the slope in \eqref{eq:R2_rose_small} is approximately
$3.550\ldots$, % This should be accurate to at least 6dp :  3.550355692
which is slightly larger than the corresponding slope for star graphs
quantised with the Laplace operator \eqref{eq:R2_star_small}.%
% Which is 2.720699...

In order to verify the calculation of the constant $c$ we have performed a
Monte-Carlo simulation. By randomly generating $10^6$ realisations of the
random variable
\begin{equation}
  \label{eq:42}
  \frac{(A_1+A_2+A_3)^{3/2}}{\sqrt{A_1A_2A_3}}
\end{equation}
we found a sample average of $6.785$ with a standard deviation
of mean of $3.296\times 10^{-3}$.

\mathversion{bold}
\subsection{Large $x$ behaviour of $R_2(x)$}
\mathversion{normal}
We base our study of the pair-correlation function at large $x$ on the
trace formula.
Starting from a smoothed version of \eqref{eq:R2_def} and applying a standard
argument, using the trace formula, we get that
\begin{align}
  \lim_{N\to\infty}\frac1{4N} \sum_{m=0}^N &\sum_{n=0}^N g(\lambda_n-
\lambda_m)  \nonumber \\ &=
%\Psi\left(\frac{\lambda_n}N\right) \Psi\left(\frac{\lambda_m}{N}\right) \\ &=
\hat{g}(0) + \frac1{\curlyL^2}\sum_{p,q\in\curlyP} \frac{A_p A_q \ell_p \ell_q}
{r_pr_q}\left(\frac{\tr d_p}2\frac{\tr d_q}2 \right)\delta_{\ell_p,\ell_q}
\left( \hat{g}\left(\frac{\ell_p}{\curlyL}\right) + \hat{g}\left(\frac{-\ell_p}
{\curlyL}\right)\right) \nonumber  \\
&= \hat{g}(0) + \int_{-\infty}^\infty \hat{g}(\tau) K_2(\tau)\,\rmd\tau,
\label{eq:form_factor1}
\end{align}
where $K_2(\tau)$ is the distribution defined by
\begin{equation}
  \label{eq:form_factor}
K_2(\tau) \coloneq \frac1{\curlyL^2}
\sum_{p,q\in\curlyP} \frac{A_p A_q \ell_p \ell_q}
{r_pr_q}\left(\frac{\tr d_p}2\frac{\tr d_q}2 \right)
\delta\left( |\tau| - \frac{\ell_p}{\curlyL} \right)
\delta_{\ell_p,\ell_q}.
\end{equation}
The quantity \eqref{eq:form_factor} is sometimes called the
\emph{pair-correlation form factor} and its behaviour as $\tau\downarrow 0$
will determine the large $x$ behaviour of $R_2(x)$.

The leading-order behaviour of the form factor can be determined using
Berry's diagonal argument \cite{ber:sto}, and was investigated for generic
graphs quantised with the Dirac operator in \cite{bol:ssf}.

Let us denote by $\gothicL$ the set of all possible lengths of
periodic orbits. Orbits with exactly the same length are said to belong
to the same \emph{degeneracy class}. We sort the sums
in \eqref{eq:form_factor} by degeneracy class:
\begin{equation}
  \label{eq:form_factor2}
K_2(\tau) = \frac1{\curlyL^2}\sum_{\ell\in\gothicL} \ell^2
\delta\left( |\tau| - \frac{\ell}{\curlyL} \right)
\bigg( \sum_{\substack{ p\in\curlyP \\ \ell_p = \ell }} \frac{A_p}{r_p}
\frac{\tr d_p}2 \bigg)^2.
\end{equation}
If we define
\begin{equation}
  \label{eq:16}
  \tilde{K}(t,B) \coloneq \frac{2B}{\curlyL^2}\sum_{\ell\in\gothicL}
\ell^2 \bigg( \sum_{\substack{ p\in\curlyP_t \\ \ell_p = \ell }} \frac{A_p}{r_p}
\frac{\tr d_p}2 \bigg)^2,
\end{equation}
where $\curlyP_t$ is the set of periodic orbits of topological length $t$ steps,
then $\tilde{K}(t,B)\to K_2(\tau)$ weakly as $B\to\infty$ provided that
$t/2B\to\tau$ as $B\to\infty$. (One proves this by integrating \eqref{eq:16} and
\eqref{eq:form_factor2} against a test function localised at $\tau$ and
noting that the bounds that we imposed on the bond lengths imply that the
delta functions corresponding to orbits with different topological lengths
are supported on disjoint intervals.)

Let us order the sum over $\ell$ according to the number of distinct
bonds to which the sum is confined, and average with respect to the random
spin matrices. Then
\begin{equation}
  \label{eq:K_tilde_expansion}
  \E( K_2(\tau) ) = \lim_{\substack{B\to\infty\\ t/2B\to\tau}} \sum_{j=1}^B
\tilde{K}_j(t,B),
\end{equation}
where
\begin{equation}
  \label{eq:K_tilde_j}
  \tilde{K}_j(t,B)\coloneq \frac{2B}{\curlyL^2}\sum_{\substack{\text{$\ell$
restricted}\\\text{to $j$ bonds}}}\ell^2
\E\bigg( \sum_{\substack{ p\in\curlyP_t \\ \ell_p = \ell }} \frac{A_p}{r_p}
\frac{\tr d_p}2 \bigg)^2,
\end{equation}
Our approximation to the averaged form factor will be based on
selecting from the sums in \eqref{eq:K_tilde_j} only those orbits
which back-scatter the maximum number of times in their degeneracy
class. As back-scattering is increasingly more strongly weighted as
$B\to\infty$ for a rose graph, this may be expected to give a good
approximation to the form factor. This type of approximation was
considered for Neumann star graphs in \cite{ber:tps}, where it was
compared to an exact expansion of the form factor, and was shown to
reproduce exactly the first four terms of the Maclaurin expansion of
the form factor.

Let us first consider the special case $j=1$ of periodic orbits confined
to a single bond $b$ of the graph. We will (unlike in \cite{ber:tps}) need
also to consider the parity of $t$.

In the case that $t$ is even, we can back-scatter $t$ times by
bouncing\footnote{We will refer to a transition of the form $b\bar{b}$ or
$\bar{b}b$ as a \textit{bounce}.}
back-and-forth along a single bond, so the contributing
orbits are $t/2$ repetitions of $b\bar{b}$ (denoting by $\bar{b}$
the reversal of $b$) with $B$ choices for the bond $b$. For such an
orbit $p$ the repetition number is $r_p=t/2$, the stability amplitude
is $A_p=(1-1/B)^t$ and $d_p = \id$. Approximating the length of
the orbit by $t$ (as each bond length approaches $1$ as $B\to\infty$)
we get
\begin{equation}
  \label{eq:15}
   \lim_{\substack{B\to\infty\\t/2B\to\tau}}\tilde{K}_1(t,B) \approx
  \lim_{\substack{B\to\infty\\t/2B\to\tau}} B \frac{2B}{\curlyL^2}t^2
\left(1-\frac1B\right)^{2t}\frac4{t^2}\left(\frac22\right)^2.
\end{equation}
As $\curlyL\to 2B$ and $t\sim2B\tau$, we get
\begin{align}
  \label{eq:17}
   \lim_{\substack{B\to\infty\\t/2B\to\tau}}\tilde{K}_1(t,B) &\approx
\lim_{B\to\infty} 2\left( 1-\frac1B \right)^{4B\tau} \nonumber \\
&=2\rme^{-4\tau},
\end{align}
for $t$ even.

For $t$ odd, we can back-scatter $t-1$ times by bouncing back-and-forth
and the final bond passed over is the same as the penultimate one; for
each of $B$ bonds, there are two orbits with maximal back-scattering,
\begin{equation}
  b\bar{b}b\bar{b} \cdots b\bar{b}\bar{b} \qquad\text{and}\qquad
\bar{b}b \bar{b}b \cdots \bar{b}bb.
\end{equation}
For these two orbits, respectively $\tr d_p = \tr w_b$ and
$\tr d_p = \tr w_{\bar{b}} = \tr w_b$, referring to
\eqref{eq:12} and \eqref{eq:w_inverse}. Both orbits have
\begin{equation}
  \label{eq:20}
  A_p = \left(\frac1{B}-1\right)^{t-1}\frac1B
\end{equation}
and $\ell_p\sim t$ and $r_p=1$. Therefore, for $t$ odd, we have
\begin{align}
  \label{eq:21}
   \lim_{\substack{B\to\infty\\t/2B\to\tau}}\tilde{K}_1(t,B) &\approx
  \lim_{\substack{B\to\infty\\t/2B\to\tau}} B \frac{2B}{\curlyL^2}t^2
\left(1-\frac1B\right)^{2t-2}\frac1{B^2}\frac14\E\left((\tr w_{\bar{b}} +
\tr w_b)^2 \right) \nonumber \\
&=\frac18 \lim_{B\to\infty} \left(1-\frac1B\right)^{4\tau B - 2} 4\tau^2
\E\left( 4 (\tr w_b)^2 \right) \nonumber\\
&=2\E(\tr w_b)^2 \tau^2 \rme^{-4\tau}.
\end{align}
For a random matrix $w\in\SU(2)$ with Haar measure, $\E(\tr w)^2=1$, so we
get
\begin{equation}
  \label{eq:22}
\lim_{\substack{B\to\infty\\t/2B\to\tau}}\tilde{K}_1(t,B) \approx
 2\tau^2\rme^{-4\tau}.
\end{equation}
Since odd and even values of $t$ have relative density $1/2$, we
 find the total contribution
for orbits confined to a single bond is the average of the two cases
in \eqref{eq:22} and \eqref{eq:17}:
\begin{equation}
  \label{eq:K_twiddle_1}
  (1+\tau^2)\rme^{-4\tau}.
\end{equation}

We next will consider the case $j=2$. This is more typical of the general case.
We now consider orbits confined to two bonds, which we will denote by
$a$ and $b$. To maximise back-scattering, we bounce $t_1$ times on bond $a$ and
$t_2$ times on bond $b$, so that $t_1+t_2=t$. In this way we can achieve
$t-2$ back-scatterings (since two transitions are necessarily between
different bonds).

For $t_1$ and $t_2$ there are four possibilities:-
\begin{alignat*}{2}
  &\left. \begin{array}{ll}
    \mbox{\bulletpoint\; $t_1$ odd, $t_2$ even} & \mbox{($\frac{t-1}2$ cases)} \\
    \mbox{\bulletpoint\; $t_1$ even, $t_2$ odd} & \mbox{($\frac{t-1}2$ cases)}
  \end{array}\right\}
  &\quad&\Rightarrow \mbox{$t$ odd,}\\
&\left. \begin{array}{ll}
  \mbox{\bulletpoint\; $t_1$ even, $t_2$ even} & \mbox{($\frac {t-2}2$ cases)} \\
   \mbox{\bulletpoint\; $t_1$ odd, $t_2$ odd} & \mbox{($\frac t2$ cases)}
  \end{array}\right\}
  &&\Rightarrow \mbox{$t$ even.}
\end{alignat*}
Since odd and even values of $t$ have relative density $1/2$, each possibility
is weighted approximately $t/4$ in its contribution to the form factor as
$t$ becomes large.

In the first of the four possibilities listed above there are four members of
the degeneracy class, which we may label as:
\begin{gather*}
  \label{eq:24}
  a\bar{a} a\bar{a} \cdots a\bar{a}a b\bar{b} \cdots b\bar{b} \\
  \bar{a}a \bar{a}a \cdots \bar{a}a\bar{a} b\bar{b} \cdots b\bar{b}  \\
  a\bar{a} a\bar{a} \cdots a\bar{a}a \bar{b}b \cdots \bar{b}b \\
  \bar{a}a \bar{a}a \cdots \bar{a}a\bar{a} \bar{b}b \cdots \bar{b}b
\end{gather*}
%For all these members we find $\tr d_p = \tr w_a$
The values of $\tr d_p$ corresponding to these four orbits reduce
to (respectively):
\begin{equation}
  \label{eq:18}
  \begin{split}
    & \tr w_a, \\
    & \tr w_{\bar a} = \tr w_a, \\
    & \tr w_a, \\
    & \tr w_{\bar a} = \tr w_a. \\
  \end{split}
\end{equation}
Similarly, to the second of the four possibilities, the we find four members
of the degeneracy class, each with $\tr d_p = \tr w_b$.

In the third possibilities, each bond is paired with its reversal an equal
number of times, and the four members of the degeneracy class have
$\tr d_p = \tr \id = 2$.

Finally, for the final possibility, the values of $\tr d_p$ are
(respectively):
\begin{equation}
  \label{eq:19}
  \begin{split}
    & \tr {w_a w_b}, \\
    & \tr {w_{\bar{a}} w_b}, \\
    & \tr {w_a w_{\bar{b}}}, \\
    & \tr {w_{\bar{a}} w_{\bar{b}}}.
  \end{split}
\end{equation}
This careful categorisation of each case reveals that the value of number
of $\mathrm{SU}(2)$ matrices appearing in the expressions for $\tr d_p$
depends only on the number of \emph{odd} values of $k_j$.

In all cases, we have:
\begin{align}
  A_p &= \left(\frac1B - 1\right)^{t-2} \frac1{B^2}, \\
  \ell_p &\sim t, \\
  r_p &= 1,
\end{align}
and there are $\displaystyle \frac{B(B-1)}2 \sim \frac{B^2}2$
choices for the pair $(a,b)$, since the $B(B-1)$ free choices of bonds
would lead to cyclic permutations, which should be counted only once
in the trace formula.

Putting the ingredients together, we have
\begin{align}
  \lim_{\substack{B\to\infty\\t/2B\to\tau}}\tilde{K}_2(t,B) &\approx
  \lim_{\substack{B\to\infty\\t/2B\to\tau}} \frac{B^2}2 \frac{2B}{\curlyL^2}
t^2 \left(1-\frac1B\right)^{2t-2}\frac1{B^4}\frac{t}4 \frac14
\bigg( \E(4^2(\tr w_a)^2) +\E(4^2(\tr w_b)^2) \nonumber \\
  \label{eq:25}
&\qquad\qquad + 8^2 + \E\left( (\tr {w_a w_b} + {w_{\bar{a}} w_b} +
{w_a w_{\bar{b}}} + {w_{\bar{a}} w_{\bar{b}}})^2\right)\bigg).
\end{align}
As before, $\E((\tr w_a)^2) = \E((\tr w_b)^2) = 1$, and we find that
\begin{equation}
  \label{eq:26}
  \E\left( (\tr {w_a w_b} + {w_{\bar{a}} w_b} +
{w_a w_{\bar{b}}} + {w_{\bar{a}} w_{\bar{b}}})^2\right) = 4,
\end{equation}
as a special case of equation \eqref{eq:trace_sum1} below. Therefore,
\begin{align}
    \lim_{\substack{B\to\infty\\t/2B\to\tau}}\tilde{K}_2(t,B) &\approx
\lim_{B\to\infty} \frac{8\tau^3}{64}\left( 1-\frac1B\right)^{4\tau B}
\left( 16+16+64+4 \right) \nonumber \\
&=\frac{25}2\tau^3\rme^{-4\tau}.
  \label{eq:K_twiddle_2}
\end{align}

We now turn to the general case of orbits confined to $j$ bonds. An orbit
with maximal back-scattering will bounce along bond $a_\ell$ a total
of $t_\ell$ times, $\ell=1,\ldots,j$, in such a way that
\begin{equation}
  \label{eq:t_sum}
  t_1 + \cdots + t_j = t.
\end{equation}
In this way, $t-j$ back-scatterings are achieved.

The weighting that each possibility receives in its contribution to the
form factor has two components. The first component is based on the
relative density of occurrences of odd or even $t_\ell$s. Since there
are  are $2^j$ possible choices for $t_\ell$ to be odd or even, this
factor is simply $1/2^j$. A second factor comes from the number of
ways to decompose $t$ in the form \eqref{eq:t_sum}. The number of
such decompositions is the number of ways to choose $j-1$ numbers
(the transition points: $t_1$, $t_1+t_2$, \textit{et cetera}) from
a total of $t-1$ possible ones.  Although these two factors are not
independent, since the odd and even choices for $t_\ell$ are evenly
distributed, the weight of each possibility is approximately their
product:
\begin{equation}
  \label{eq:23}
  \frac1{2^j}\binom{t-1}{j-1} \sim \frac{t^{j-1}}{2^j (j-1)!}.
\end{equation}

% There are $2^j$ possible choices for $t_\ell$ to be odd or even. Each
% possibility will be weighted approximately
% \begin{equation}
%   \label{eq:23}
%   \frac1{2^j}\binom{t-1}{j-1} \sim \frac{t^{j-1}}{2^j (j-1)!}
% \end{equation}
% in its contribution to the form factor (the binomial coefficient is the
% number of degeneracy classes).

Within each possibility it is the number of odd $t_j$s that determines the
value of the trace factor; after cancellations we are left with
\begin{equation}
  \label{eq:trace_contribution}
  \tr d_p = \tr\left( w_{a_{i_1}}^{\alpha_{i_1}} \cdots
w_{a_{i_r}}^{\alpha_{i_r}} \right),
\end{equation}
where there are precisely $r$ indices $\{ i_1,\ldots,i_r \}$ for which
$t_{i_{\ell}}$ is odd, and each $\alpha_{i_\ell}=\pm1$. There are $2^j$
members of each degeneracy class, and $2^r$ ways that $\alpha_{i_\ell}=\pm1$.
So we will need to calculate
\begin{equation}
  \label{eq:expectation_need}
  \E\Bigg\{ \Bigg( 2^{j-r} \sum_{\alpha_{i_\ell}=\pm 1} \tr\left(
w_{a_{i_1}}^{\alpha_{i_1}} \cdots w_{a_{i_r}}^{\alpha_{i_r}} \right)\Bigg)^2
\Bigg\}.
\end{equation}
Since every combination of $\alpha_{i_\ell}=\pm1$ occurs in the sum, and
$w_{a_{i_\ell}}+w_{a_{i_\ell}}^{-1} = \tr(w_{a_{i_\ell}})\id$
we have
\begin{equation}
  \label{eq:trace_sum0}
   \sum_{\alpha_{i_\ell}=\pm 1}
w_{a_{i_1}}^{\alpha_{i_1}} \cdots w_{a_{i_r}}^{\alpha_{i_r}} =
(\tr w_{a_1})\cdots(\tr w_{a_r})\id.
\end{equation}
Thus,
\begin{align}
    \E\Bigg\{ \Bigg( 2^{j-r} \sum_{\alpha_{i_\ell}=\pm 1} \tr\left(
w_{a_{i_1}}^{\alpha_{i_1}} \cdots w_{a_{i_r}}^{\alpha_{i_r}} \right)\Bigg)^2
\Bigg\} &= 4^{j+1-r} \E\Big\{ (\tr w_{a_1})^2\cdots(\tr w_{a_r})^2 \Big\}
\nonumber \\
&=4^{j+1-r}.
  \label{eq:trace_sum1}
\end{align}
Finally, we note that there are $\binom{j}{r}$ combinations of ways that there
can be $r$ odd indices out of a total of $j$.

In all cases, we have:
\begin{align}
  A_p &= \left(\frac1B - 1\right)^{t-j} \frac1{B^j}, \\
  \ell_p &\sim t \qquad\text{and}\\
  r_p &= 1.
\end{align}
The number of choices for the $j$ bonds on which the orbits are confined is
$\displaystyle \frac1j \frac{B!}{(B-j)!} \sim \frac{B^j}{j}$
(the factor $1/j$ is to account for cyclic invariance).
We get
\begin{align}
  \lim_{\substack{B\to\infty\\t/2B\to\tau}}\tilde{K}_j(t,B) &=
  \lim_{\substack{B\to\infty\\t/2B\to\tau}} \frac{B^j}{j} \frac{2B}{\curlyL^2}
t^2 \left(1-\frac1B\right)^{2t-2j}\frac1{B^{2j}}\frac{t^{j-1}}{2^j}
\frac{1}{(j-1)!} \frac14 \Bigg\{\sum_{j=0}^r \binom{j}{r} 4^{j+1-r} \Bigg\}
\nonumber\\
  &=\lim_{\substack{B\to\infty\\t/2B\to\tau}} \frac{t^{j+1}}{2^{j+1}B^{j+1}}
\left( 1-\frac1B \right)^{2t-2j} \frac{1}{j!}
\Bigg\{\sum_{j=0}^r \binom{j}{r} 4^{j-r} \Bigg\} \nonumber \\
 &=\tau^{j+1} \rme^{-4\tau} \frac{5^j}{j!}.
\label{eq:K_twiddle_j}
\end{align}
We remark that upon substituting $j=2$ into \eqref{eq:K_twiddle_j} we
recover \eqref{eq:K_twiddle_2}, as expected.

Substituting \eqref{eq:K_twiddle_1} and \eqref{eq:K_twiddle_j} into
\eqref{eq:K_tilde_expansion} we get
\begin{align}
\E(K_2(\tau)) &\approx (1+\tau^2)\rme^{-4\tau}  + \sum_{j=2}^\infty
\tau^{j+1} \rme^{-4\tau} \frac{5^j}{j!} \nonumber \\
 &=(1+\tau^2)\rme^{-4\tau} + \tau {\rme^{-4\tau}}\left(
\rme^{5\tau}-1-5\tau\right) \nonumber \\
  &=(1-\tau-4\tau^2)\rme^{-4\tau} + \tau\rme^\tau.
  \label{eq:K_tilde_approx}
\end{align}
Expanding \eqref{eq:K_tilde_approx} as a Maclaurin series, we get
\begin{equation}
  \label{eq:K_tilde_approx_expand}
  \E(K_2(\tau))\approx 1 - 4\tau +9\tau^2 -\frac{13}6\tau^3 +
\Ord(\tau^4),
\end{equation}
for small values of $\tau$.

It follows from the fact that the form-factor and the pair-correlation
function are related \textit{via} the Fourier transform that the small
$\tau$ asymptotics \eqref{eq:K_tilde_approx} determine the large $x$
behaviour of the averaged pair-correlation function. Namely, we have
\cite[page 102]{ber:phd} that if $k(\tau)$ is even and
\begin{equation}
  \label{eq:27}
  k(\tau)\sim 1 + \sum_{k=1}^\infty a_k\tau^k
\end{equation}
and
\begin{equation}
  \label{eq:29}
  1-k(\tau) = \int_{-\infty}^\infty (1-r(x))\rme^{2\pi\rmi x \tau}\,\rmd x,
\end{equation}
then
\begin{equation}
  \label{eq:30}
  r(x) \sim 1 + 2\Re\Bigg\{ \sum_{k=1}^\infty \left( \frac{-\rmi}{2\pi}
\right)^{k+1} \frac{a_k k!}{x^{k+1}}\Bigg\}.
\end{equation}
Applying this to the form-factor approximation
\eqref{eq:K_tilde_approx_expand}, we get the following approximation
to the pair-correlation function:
\begin{equation}
  \label{eq:R_2_approx_big}
  \E(R_2(x)) \approx 1 + \frac2{\pi^2 x^2} - \frac{13}{8\pi^4 x^4}
+ \Ord\!\left( \frac1{x^6} \right),
\end{equation}
for large values of $x$.
\section{Conclusions}\label{sec:conclusions}
We have investigated the spectral statistics of a quantum graph quantised
with the Dirac operator, for which the Schr\"odinger-operator-quantised
counterpart (Laplace operator)
has intermediate spectral statistics. The shape of the
graph has a single central vertex, and all bonds are connected at both
ends to that vertex. We call this graph a rose.

We have shown that the generic condition for $k$ to be an eigenvalue is that it
satisfies the non-linear equation
\begin{equation}
  \label{eq:eigenvalue_condition_con}
 \sum_{b=1}^B \frac{\cos\theta_b - \cos kL_b}{\sin kL_b} = 0,
\end{equation}
where the angles $\theta_b$ are determined by a set of $B$ matrices
from $\mathrm{SU}(2)$ that rotate the spinor during its passage along
the bonds. Each eigenvalue occurs with multiplicity $2$ (Kramer's degeneracy).

We have investigated the behaviour of the spectral pair correlation
function $R_2(x)$ in the limit $B\to\infty$ for large and small values
of $x$, when it is averaged over random realisations of the
$\mathrm{SU}(2)$ matrices chosen with Haar measure.

For small values of $x$ we have found
\begin{equation}
  \label{eq:R2_rose_small_con}
 \E( R_2(x)) \approx \frac{\pi c}{6}x \qquad\text{as $x\to0$,}
\end{equation}
where  $c \approx 6.781\ldots$

For large values of $x$ we have determined
\begin{equation}
  \label{eq:R2_rose_big_con}
  \E(R_2(x)) \approx 1 + \frac2{\pi^2 x^2} - \frac{13}{8\pi^4 x^4}
+ \Ord\!\left( \frac1{x^6} \right).
\end{equation}

The behaviour of the pair correlation function is different to the pair
correlation function for star graphs quantised with the Laplace operator
(see also figure \ref{fig:tre} below for a numerical comparison). It
is too early to conjecture that the behaviour that we found is universal
for systems with intermediate statistics and a symplectic symmetry, but
this possibility merits further investigation.

It is interesting to note that while we considered elements $w_b$ chosen with Haar measure in $\SU (2)$ the large parameter asymptotic of $R_2(x)$ would be the same if $w_b$ is chosen from any irreducible representation of a subgroup $\Gamma\subset\SU (2)$.  The reason for this was seen in (\ref{eq:trace_sum1}) where to evaluate the asymptotic it was only necessary to know that $\E\{  (\tr w)^2\}=1$.  As $\tr w$ is the character of an element of the subgroup when the average is carried out over $\Gamma$ for an irreducible representation of $\Gamma$ the result must still be one by the character orthogonality relations.  So, for example, choosing the spin transformations $w_b$ on the rose from the finite subgroup
\begin{equation}\label{eq:Q8}
\Gamma=\{\pm \UI, \pm\ui \sigma_x, \pm\ui \sigma_y, \pm \ui \sigma_z \}  \ ,
\end{equation}
where $\sigma_j$ is a Pauli matrix, will not change the large $x$ asymptotic of $R_2(x)$.
The small parameter asymptotic, in contrast, depends on the distribution of the $A_j$ given in (\ref{eq:A_density}) which will vary if the spin transformations are chosen from an irreducible representation of a subgroup.

\subsection{Some numerics}
In order to numerically verify the small and large $x$ behaviour of
the pair correlation function found in the previous section, we
performed numerical calculations of the eigenvalues of rose graphs
quantised with the Dirac operator, and calculated the empirical pair
correlation statistic. To implement the averaging over the random choice
of  $\mathrm{SU}(2)$ matrices at the graph vertex, we performed the calculations
100 times with random Haar-distributed matrices and averaged the results.
In each realisation, 150\,000 eigenvalues were calculated.

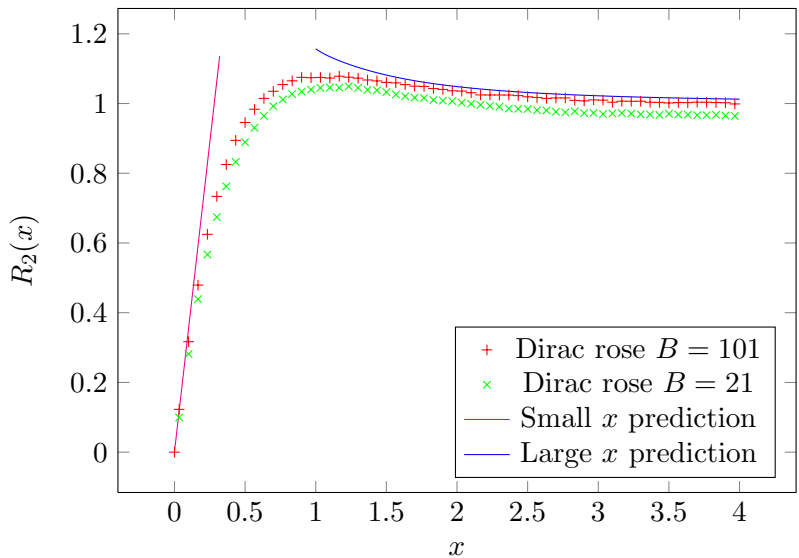
\begin{figure}[h]
\begin{center}
  \begin{tikzpicture}
    \begin{axis}[
      xlabel=$x$,
      ylabel=$R_2(x)$]
      \addplot[color=red,
               mark=+,
               only marks] file {rose100-pcf-101-150K.dat};
      \addlegendentry{Dirac rose $B=101$}
      \addplot[color=green,
               mark=x,
               only marks] file {rose100-pcf-21-150K.dat};
      \addlegendentry{Dirac rose $B=21$}
      \addplot[color=magenta, no markers]
               coordinates {(0,0) (0.32,1.136113821)};
      \addlegendentry{Small $x$ prediction}
      \addplot[color=blue, no markers] file {rosepaircorr-big.dat};
      \addlegendentry{Large $x$ prediction}
    \end{axis}
  \end{tikzpicture}
\caption{The pair-correlation function numerically calculated for a $B=21$ bond
(green) and a $B=101$ bond (red) Dirac rose graph.
Also plotted are the curves corresponding to the
large and small values of the parameter, \eqref{eq:R2_rose_big_con} and
\eqref{eq:R2_rose_small_con} respectively. The numerical plots are averaged over
100 realisations of the $\mathrm{SU}(2)$ matrices, and 150\,000 eigenvalues
were computed.}
\label{fig:due}
\end{center}
\end{figure}

In figure \ref{fig:due} we compare the numerically-calculated pair
correlation function with the predictions of \eqref{eq:R2_rose_small_con} and
\eqref{eq:R2_rose_big_con}, and find a good agreement in the range of
validity.

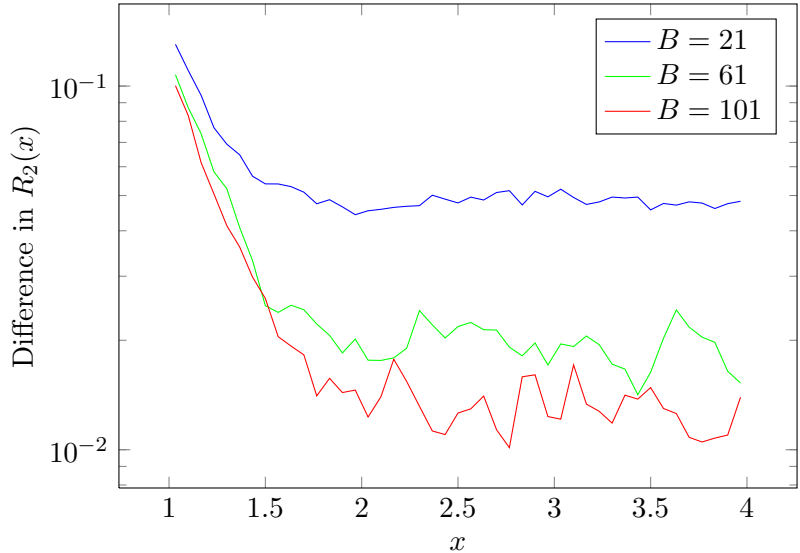
\begin{figure}[h]
\begin{center}
  \begin{tikzpicture}
    \begin{semilogyaxis}[
      xlabel=$x$,
      ylabel={Difference in $R_2(x)$},
      legend style={
        legend pos=north east,
        cells={anchor=west}}]
      \addplot[color=blue, no markers] table[x index=0, y index=3]
                      {cf-rose100-pcf-101-61-21-150K.dat};
      \addlegendentry{$B=21$}
      \addplot[color=green, no markers] table[x index=0, y index=2]
                      {cf-rose100-pcf-101-61-21-150K.dat};
      \addlegendentry{$B=61$}
      \addplot[color=red, no markers] table[x index=0, y index=1]
                      {cf-rose100-pcf-101-61-21-150K.dat};
      \addlegendentry{$B=101$}
    \end{semilogyaxis}
  \end{tikzpicture}
\caption{The difference between the asymptotic \eqref{eq:R2_rose_big_con}
and the pair-correlation function numerically calculated for a $B=21$ bond
(blue), $B=61$ bond (green) and $B=101$ bond (red)
Dirac rose graph, plotted on a logarithmic scale.}
%150\,000 eigenvalues were computed.}
\label{fig:cinque}
\end{center}
\end{figure}

The agreement gets better as the number of bonds increases, as is to
be expected since our analytical calculations relate to the limit
$B\to\infty$. This improvement is demonstrated in figure
\ref{fig:cinque} in which a comparison is made between the
pair-correlation function for 21, 61 and 101 bond graphs, and the
large $x$ prediction of \eqref{eq:R2_rose_big_con}. A clear increase
in adherence to the prediction is displayed as the number of bonds increases.

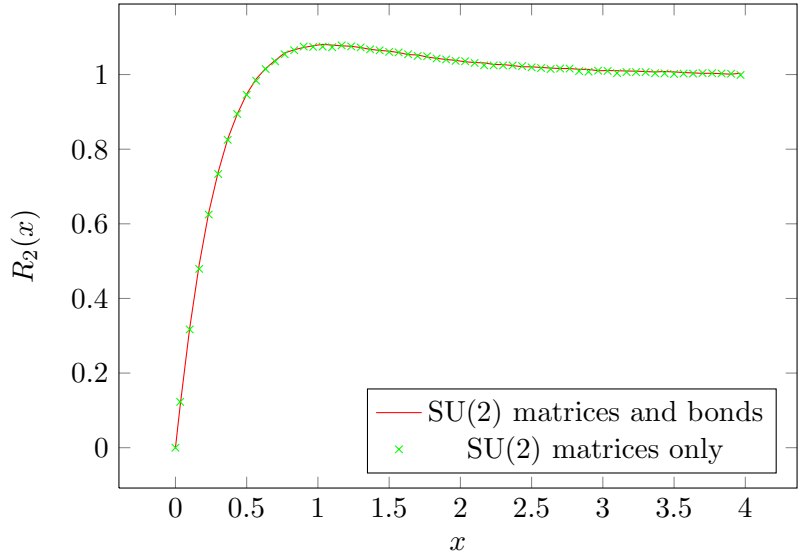
\begin{figure}[h]
\begin{center}
  \begin{tikzpicture}
    \begin{axis}[
      xlabel=$x$,
      ylabel=$R_2(x)$]
      \addplot[color=red, no markers] file {rRose100-pcf-101-150K.dat};
      \addlegendentry{$\mathrm{SU}(2)$ matrices and bonds}
      \addplot[color=green,
               mark=x,
               only marks] file {rose100-pcf-101-150K.dat};
      \addlegendentry{$\mathrm{SU}(2)$ matrices only}
    \end{axis}
  \end{tikzpicture}
\caption{The pair-correlation function numerically calculated for a $B=101$
bond Dirac rose graph (green),
100 realisations of the $\mathrm{SU}(2)$ matrices only (green points) and
100 realisations of the matrices \emph{and} bond lengths (red curve).
150\,000 eigenvalues were computed.}
\label{fig:quattro}
\end{center}
\end{figure}

One may wonder if averaging over both the random $\mathrm{SU}(2)$ matrices
and bond lengths will lead to a different pair-correlation function. Figure
\ref{fig:quattro} compares such a calculated empirical curve with that for
corresponding graph with averaging over the $\mathrm{SU}(2)$ matrices
only, for a $B=101$ bond Dirac rose graph. There is no noticeable difference
in the curves so obtained.

\begin{figure}[h]
\begin{center}
  \begin{tikzpicture}
    \begin{axis}[
      xlabel=$x$,
      ylabel=$R_2(x)$]
      \addplot[color=green, no markers] file {rose100-pcf-101-150K.dat};
      \addlegendentry{Dirac rose}
      \addplot[color=red, no markers] file {star100-pcf-101-150K.dat};
      \addlegendentry{Neumann star}
    \end{axis}
  \end{tikzpicture}
\caption{A comparison of the pair-correlation function numerically calculated
for a $B=101$ bond Neumann star graph (red) and Dirac rose graph (green).
The numerical plots are averaged over
100 realisations of the $\mathrm{SU}(2)$ matrices for the rose graph and
100 realisations of the bond lengths for the star graph, and 150\,000
eigenvalues were computed.}
\label{fig:tre}
\end{center}
\end{figure}
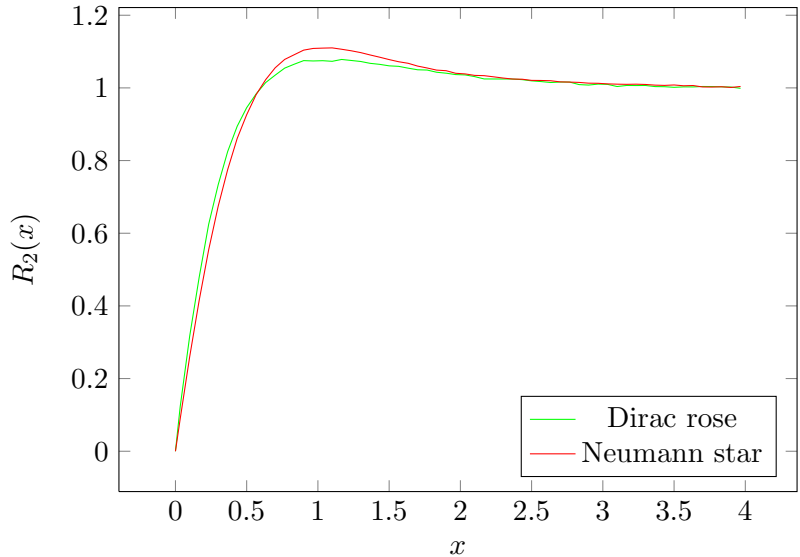

It is of interest to compare the pair correlation function for the Neumann star
graphs with the Laplace operator, and the Dirac rose graphs. This is done in
figure \ref{fig:tre} for graphs with 101 bonds in both cases. Qualitatively the
curves are similar in appearance, but there is some noticeable difference,
which can already be explained analytically around the point $x=0$
(compare \eqref{eq:R2_rose_small_con} and \eqref{eq:R2_star_small}).

\subsection*{Acknowledgements}
We are grateful to
\href{http://www.maths.bris.ac.uk/people/profile/majpk}{Jon Keating}
for encouraging us to work on this problem,
and acknowledge fruitful conversations with
\href{http://www.math.tamu.edu/~berko/}{Gregory Berkolaiko},
and
\href{http://www.ma.rhul.ac.uk/jbolte}{Jens Bolte}
regarding this work.

JMH would like to thank Bristol University for their hospitality
during his sabbatical during which some of the work was carried out.
BW has been financially supported by EPSRC grant number EP/H046240/1.
JMH was supported by the Baylor University research leave and summer
sabbatical programs.

% rose100-pcf-101-21-150K.pdf
%
% rose-star100-pcf-101-150K.pdf
%
% rNrRose100-pcf-101-150K.pdf
%
% cf-ll-rose100-pcf-101-62-21-150K.pdf
%
%\bibliographystyle{brianbib2}
%\bibliography{references,unpublished}
%
%%%%%%%%%%%%%%%%%%%%%%%%%%%%%% Bibliography %%%%%%%%%%%%%%%%%%%%%%%%%%%%%
%

\end{document}